\documentclass[11pt]{article}
\usepackage[totalwidth=460truept,totalheight=600truept]{geometry}
\usepackage{epsfig,latexsym,amssymb}

\def\theequation{\arabic{section}.\arabic{equation}}

\renewcommand{\theequation}{\thesection.\arabic{equation}}
\global\arraycolsep=1truept
\input epsf
\input{tcilatex}

\begin{document}

\hfill \hfill IFUP-TH 2005/22

\vskip 1.4truecm

\begin{center}
{\huge \textbf{Dimensionally Continued Infinite Reduction }}

\vskip .4truecm

{\huge \textbf{Of Couplings}}

\vskip 1.5truecm

\textsl{Damiano Anselmi and Milenko Halat}

\textit{Dipartimento di Fisica ``Enrico Fermi'', Universit\`{a} di Pisa, }

\textit{Largo Bruno Pontecorvo 3, I-56127 Pisa, Italy, }

\textit{and INFN, Sezione di Pisa, Italy}

anselmi@df.unipi.it, halat@df.unipi.it
\end{center}

\vskip 2truecm

\begin{center}
\textbf{Abstract}
\end{center}

{\small The infinite reduction of couplings is a tool to consistently
renormalize a wide class of non-renormalizable theories with a reduced,
eventually finite, set of independent couplings, and classify the
non-renormalizable interactions. Several properties of the reduction of
couplings, both in renormalizable and non-renormalizable theories, can be
better appreciated working at the regularized level, using the
dimensional-regularization technique. We show that, when suitable
invertibility conditions are fulfilled, the reduction follows uniquely from
the requirement that both the bare and renormalized reduction relations be
analytic in }$\varepsilon =D-d${\small , where }$D${\small \ and }$d${\small %
\ are the physical and continued spacetime dimensions, respectively. In
practice, physically independent interactions are distinguished by
relatively non-integer powers of }$\varepsilon ${\small . We discuss the
main physical and mathematical properties of this criterion for the
reduction and compare it with other equivalent criteria. The leading-log
approximation is solved explicitly and contains sufficient information for
the existence and uniqueness of the reduction to all orders.}

{\small \ }

\vskip1truecm

\vfill\eject

\section{Introduction}

\setcounter{equation}{0}

Formulated in the common fashion, non-renormalizable theories can be used
only as low-energy effective field theories, due to the presence of
infinitely many independent couplings, introduced to subtract the
divergences. Although effective field theories are good for most practical
purposes, they are unable to suggest new physics beyond them. The search for
more fundamental theories that include non-renormalizable interactions can
have applications both to physics beyond the Standard Model and quantum
gravity. Moreover, it can clarify if quantum field theory is really
inadequate, as some physicists think, to explain the fundamental
interactions of nature beyond power-counting renormalizability. In that
case, it can suggest a more fundamental theoretical framework to supersede
quantum field theory.

The main tool to classify the irrelevant interactions is the ``infinite
reduction'', that is to say the reduction of couplings in non-renormalizable
theories. The reduction of couplings was first studied by Zimmermann \cite
{zimme,oheme} for power-counting renormalizable theories, as an alternative
to unification. The idea is to look for relations among the coupling
constants that are consistent with the renormalization of divergences. Such
relations, rather than being based on symmetries, follow from\ constraints
on the solutions to the renormalization-group (RG) equations. The constraint
proposed by Zimmermann is that the reduction should be analytic. Assume that
a theory containing two couplings $\alpha $ and $\lambda $ is an effective
formulation of a more fundamental theory containing only the coupling $%
\alpha $. It is clear that if the connection between the more fundamental
theory and the effective theory is perturbative, then the relation $\lambda
(\alpha )$ should be analytic in $\alpha $, because a perturbative expansion
cannot generate fractional, irrational, complex or negative-integer powers
of $\alpha $.

A non-renormalizable interaction is made of infinitely many lagrangian
terms. The infinite reduction is the search for special relations that
express the irrelevant couplings as unique functions of a reduced,
eventually finite, set of independent couplings $\overline{\lambda }$, such
that the divergences are removed by means of field redefinitions plus
renormalization constants for the $\overline{\lambda }$s. It can be easily
shown that in non-renormalizable theories the analyticity requirement
proposed by Zimmermann is too strong. The basic reason is that an irrelevant
deformation has no end, so it is impossible to identify the minimum power of
a coupling, which is necessary to demand analyticity. Then it is natural to
allow also for negative powers of the couplings and replace analyticity by
perturbative meromorphy. That means that the reduction relations have to be
meromorphic in the marginal couplings $\alpha $, and that the maximum
negative $\alpha $-powers should be bounded by the order of the perturbative
expansion \cite{infredfin}.

Although perturbative meromorphy is an exhaustive criterion for the infinite
reduction, it is useful to build a framework where the reduction is more
natural and its properties can be more clearly appreciated. For this
purpose, it is convenient to study the reduction of couplings at the
regularized level, using the dimensional-regularization technique. The
reduction of couplings is by construction RG\ invariant, so in a common
regularization framework the bare and renormalized reduction relations look
the same, because the bare couplings are nothing but the renormalized
couplings calculated at an energy scale equal to the cut-off. However, the
dimensional-regularization technique makes systematic use of the parameter $%
\varepsilon =D-d$, $D$ and $d$ being the physical and continued spacetime
dimensions, respectively, and various properties can be conveniently
formulated in terms of analyticity and meromorphy in the parameter $%
\varepsilon $.

\bigskip

We give two equivalent criteria for the infinite reduction. The first
criterion is obtained studying the analyticity properties of the
renormalized reduction relations and is an immediate generalization of the
criterion formulated in ref. \cite{infredfin} at $\varepsilon =0$. The
second criterion is new, obtained comparing the analyticity properties of
the renormalized and bare reduction relations.

If the theory contains a unique strictly-renormalizable coupling $\alpha $,
when certain invertibility conditions hold, the reduction is uniquely
determined by the following two equivalent requirements:

\noindent 1) that the renormalized reduction relations be perturbatively
meromorphic in $\alpha $ at $\alpha =0$ and analytic in $\varepsilon $ at $%
\varepsilon =0$;

\noindent \noindent 2) that the renormalized and bare reduction relations be
analytic in $\varepsilon $ at $\varepsilon =0$.

Requirement 2) will be called \textit{double analyticity in }$\varepsilon $.
Suitable generalizations of the criteria just stated apply to the theories
that contain more marginal couplings.

Because of meromorphy in $\alpha $, the strictly-renormalizable subsector of
the theory must be fully interacting.

\bigskip

The non-renormalizable interactions can be studied expanding the correlation
functions perturbatively in the overall energy $E$, for $E\ll M_{P\mathrm{eff%
}}$, where $M_{P\mathrm{eff}}$ is an ``effective Planck mass''.

It is important to emphasize that in our analysis all super-renormalizable
parameters, in particular the masses, are switched off, which is consistent
in appropriate subtraction schemes, including the dimensional-regularization
technique. This assumption ensures that the beta functions of the
non-renormalizable couplings are polynomial in the non-renormalizable
couplings themselves allowing an explicit solution of the reduction
equations by a simple recursive procedure. The super-renormalizable
parameters can be eventually incorporated in a second stage perturbatively
in the reduction equations.

\bigskip

Now we illustrate the main features of the infinite reduction at the
regularized level. Let $\alpha $ and $\lambda $ denote two independent
couplings. In general, the bare and renormalized reduction relations have
the form 
\begin{equation}
\lambda _{\mathrm{B}}=\lambda _{\mathrm{B}}(\alpha _{\mathrm{B}},\zeta
,\varepsilon ),\qquad \lambda =\lambda (\alpha ,\xi ,\varepsilon ),\noindent
\label{reno}
\end{equation}
respectively, where $\zeta $ and $\xi $ are certain arbitrary constants,
defined so that (\ref{reno}) are analytic in $\varepsilon $, $\zeta $ and $%
\xi $. Matching the relations (\ref{reno}) with each other, formulas 
\begin{equation}
\zeta =\zeta (\xi ,\varepsilon ),\qquad \Leftrightarrow \qquad \xi =\xi
(\zeta ,\varepsilon )  \label{zx}
\end{equation}
relating the constants $\zeta $ and $\xi $ can be worked out. In general,
the relations (\ref{zx}) are not analytic in $\varepsilon $ at $\varepsilon
=0$, which implies that the bare and renormalized relations are not
contemporarily analytic in $\varepsilon $, for generic values of $\zeta $
and $\xi $. On the other hand, when certain invertibility conditions are
fulfilled, there exists unique values $\overline{\zeta }$ and $\overline{\xi 
}$ such that both $\lambda _{\mathrm{B}}=\lambda _{\mathrm{B}}(\alpha _{%
\mathrm{B}},\overline{\zeta },\varepsilon )$ and $\lambda =\lambda (\alpha ,%
\overline{\xi },\varepsilon )$ are analytic in $\varepsilon $. Those values
select the infinite reduction.

Independent irrelevant interactions are distinguished by relatively
fractional, irrational, complex or negative-integer powers of $\varepsilon $%
. These properties provide a criterion to classify the non-renormalizable
interactions.

The criterion just formulated can be justified as follows. The quantum
action $\Gamma [\Phi ,\alpha ,\lambda ,\varepsilon ]$ is convergent in the
physical limit, that is the limit $\varepsilon \rightarrow 0$ at fixed
renormalized fields and couplings. If a reduction $\lambda \left( \alpha
,\varepsilon \right) $ is consistent, then also the reduced quantum action $%
\Gamma [\Phi ,\alpha ,\lambda (\alpha ,\varepsilon ),\varepsilon ]$ should
be convergent in the physical limit, so $\lambda \left( \alpha ,\varepsilon
\right) $ should be regular for $\varepsilon \sim 0$. The bare lagrangian $%
\mathcal{L}(\varphi _{\mathrm{B}},\lambda _{\mathrm{B}},\alpha _{\mathrm{B}%
},\varepsilon )$, on the other hand, tends to the classical lagrangian in
the ``naive'' limit, that is the limit $\varepsilon \rightarrow 0$ at fixed
bare fields and couplings. If a reduction $\lambda _{\mathrm{B}}\left(
\alpha _{\mathrm{B}},\varepsilon \right) $ is consistent, then the reduced
bare lagrangian $\mathcal{L}(\varphi _{\mathrm{B}},\alpha _{\mathrm{B}%
},\lambda _{\mathrm{B}}(\alpha _{\mathrm{B}},\varepsilon ),\varepsilon )$
should converge to the reduced classical lagrangian in the naive limit, so $%
\lambda _{\mathrm{B}}(\alpha _{\mathrm{B}},\varepsilon )$ should be regular
for $\varepsilon \sim 0$. In perturbation theory it is safe to replace the
words ``convergent'' and ``regular'' with the word ``analytic''. Indeed, the
more fundamental theory is certainly analytic in $\varepsilon $ and if the
connection between the more fundamental theory and the effective theory is
perturbative the bare and renormalized relations should be both analytic in $%
\varepsilon $.

In ref. \cite{infredfin} the infinite reduction was studied at $\varepsilon
=0$, where the relations among the couplings are uniquely selected by
perturbative meromorphy. We will show that the infinite reduction at $%
\varepsilon \neq 0$, implied by the double $\varepsilon $-analyticity, gives
results that are physically consistent with those of ref. \cite{infredfin}.
The infinite reduction at $\varepsilon \neq 0$ has already been studied in 
\cite{renscal} for a special class of models. It is worth to recall that
when the renormalizable sector is an interacting conformal field theory the
infinite reduction has peculiar features that make it conceptually simpler 
\cite{pap2}.

\bigskip

The study of quantum field theory beyond power counting has been attracting
interest for a long time, motivated by effective field theory, low-energy
QCD, quantum gravity and the search for new physics beyond the Standard
Model. Here we are not concerned with the predictiveness of our theories.
Our purpose is merely to give mathematical tools to organize and classify
the non-renormalizable interactions, using analyticity properties and
consistency with the renormalization group.

Sometimes \textit{ad hoc} subtractions are used to make non-renormalizable
theories predictive. Often they amount to assign preferred values
(typically, zero) to an infinity of renormalized couplings at the
subtraction point $\mu $. Another example of ad hoc prescription is the BPZH
subtraction \cite{bphz} of divergent diagrams at zero momentum. These
prescriptions, although appealing for a variety of reasons, are not
consistent with the renormalization group. Our construction, on the other
hand, follows precisely from consistency with the renormalization group.

A different approach to reduce the number of independent couplings in
non-renormalizable theories is Weinberg's asymptotic safety \cite{wein},
which has been recently studied using the exact renormalization-group
techniques \cite{reuter,wette,litim}. Other investigations of reductions of
couplings in non-renormalizable theories have been performed by Atance and
Cortes \cite{cortes1,cortes2}, Kubo and Nunami \cite{giap}, Halpern and
Huang \cite{halpern}. For a recent perturbative renormalization-group
approach to non-renormalizable theories, see \cite{kazakov}.

\medskip

The paper is organized as follows. In section 2 we\ study Zimmermann's
reduction of couplings at $\varepsilon \neq 0$ and exhibit an interesting
connection with the $\varepsilon $-expansion techniques. In sections 3 and 4
we\ study the reduction in non-renormalizable theories, first at the
regularized level and then at the bare level. In section 5 we study the
physical invertibility conditions in the absence of three-leg marginal
vertices, while in section 6 we explicitly solve the infinite reduction in
the leading-log approximation, which contains enough information about the
existence and uniqueness of the reduction to all orders. In most of the
paper we assume that the renormalizable subsector contains a single
strictly-renormalizable coupling. In section 7 we generalize our results to
theories whose renormalizable subsector contains more independent couplings.
Section 8 contains the conclusions, while the appendix contains the
derivation of multivariable renormalization constants from the associated
beta functions, which is used in the paper. We work in the Euclidean
framework.

\section{Zimmermann's reduction at $\varepsilon \neq 0$}

\setcounter{equation}{0}

Consider a renormalizable theory with two marginal couplings, $\rho $ and $g$%
, such as massless scalar electrodynamics, 
\begin{equation}
\mathcal{L}=\frac{1}{4}F_{\mu \nu }^{2}+|D_{\mu }\varphi |^{2}+\frac{\rho }{4%
}(\overline{\varphi }\varphi )^{2},  \label{nonsusy}
\end{equation}
where $D_{\mu }\varphi =\partial _{\mu }\varphi +igA_{\mu }\varphi $, or the
massless Yukawa model 
\[
\mathcal{L}=\frac{1}{2}(\partial \varphi )^{2}+\overline{\psi }\partial
\!\!\!\slash \psi +g\varphi \overline{\psi }\psi +\frac{\rho }{4!}\varphi
^{4}. 
\]
Define $\alpha =g^{2}$. Write the bare couplings and renormalization
constants as 
\[
\alpha _{\mathrm{B}}=\mu ^{\varepsilon }\alpha Z_{\alpha }^{\prime }(\alpha
,\rho ,\varepsilon ),\qquad \rho _{\mathrm{B}}=\mu ^{\varepsilon }\left(
\rho +\Delta _{\rho }(\alpha ,\rho ,\varepsilon )\right) , 
\]
where $Z_{\alpha }^{\prime }$ and $\Delta _{\rho }$ are analytic functions
of the couplings. It is convenient to write $\rho =\alpha \eta $ and use $%
\alpha $ and $\eta $ as independent couplings. Then a Feynman diagram
carries a power of $\alpha $ equal to $v_{4}+v_{3}/2$, where $v_{3}$, $v_{4}$
denote the numbers of three-leg and four-leg vertices, respectively. Using $%
4v_{4}+3v_{3}=E+2I$ and $V=v_{4}+v_{3}=I-L+1$, where $E$, $L$, $I$ and $V$
denote the numbers of external legs, loops, internal legs and vertices,
respectively, we have 
\begin{equation}
v_{4}+\frac{v_{3}}{2}=\frac{E}{2}-1+L.  \label{ah}
\end{equation}
Therefore, the counterterms that renormalize the three-leg vertex are
proportional to $g\alpha^{L} $ and those that renormalize the four-leg
vertex are proportional to $\alpha ^{L+1}$. Then it is possible to write 
\begin{equation}
\alpha _{\mathrm{B}}=\mu ^{\varepsilon }\alpha Z_{\alpha }(\alpha ,\eta
,\varepsilon ),\qquad \eta _{\mathrm{B}}=\eta +\alpha \Delta _{\eta }(\alpha
,\eta ,\varepsilon ),  \label{abbare}
\end{equation}
and $Z_{\alpha }$, $\Delta _{\eta }$ are analytic functions of $\alpha $, $%
\eta $.

\bigskip

\textbf{Reduction}. The reduction is a relation $\eta (\alpha )$ between the
two couplings, such that a single renormalization constant, the one of $%
\alpha $, is sufficient to remove the divergences associated with both $%
\alpha $ and $\eta $. This goal can be achieved imposing consistency with
the renormalization group. In the minimal subtraction scheme we have 
\begin{equation}
\frac{\mathrm{d}\eta }{\mathrm{d}\ln \mu }=\beta _{\eta }(\eta ,\alpha
),\qquad \frac{\mathrm{d}\alpha }{\mathrm{d}\ln \mu }=\beta _{\alpha }(\eta
,\alpha )-\varepsilon \alpha .  \label{betas}
\end{equation}
Because of (\ref{ah}) and (\ref{abbare}), the beta functions have expansions 
\begin{equation}
\beta _{\alpha }=\beta _{1}\alpha ^{2}+\sum_{k=2}^{\infty }\alpha
^{k+1}P_{k}(\eta ),\qquad \beta _{\eta }=\alpha \left( a+b\eta +c\eta
^{2}\right) +\sum_{k=2}^{\infty }\alpha ^{k}Q_{k+1}(\eta ),  \label{abbe}
\end{equation}
where $P_{k}(\eta )$ and $Q_{k}(\eta )$ are polynomials of order $k$. The
one-loop structure of $\beta _{\alpha }$ follows from explicit analysis of
the one-loop diagrams. Consistency with the RG\ equations implies the
differential equation 
\begin{equation}
\frac{\mathrm{d}\eta (\alpha )}{\mathrm{d}\alpha }=\frac{\beta _{\eta }(\eta
(\alpha ),\alpha )}{\beta _{\alpha }(\eta (\alpha ),\alpha )-\varepsilon
\alpha },  \label{zimmesp}
\end{equation}
that determines the solution $\eta (\alpha )$ up to an arbitrary constant $%
\xi $, the initial condition. If $\beta _{\alpha }\neq 0$ the solution has a
smooth limit for $\varepsilon \rightarrow 0$, which is the Zimmermann
solution \cite{zimme}.

\bigskip

\textbf{Reduction at the level of bare couplings}. The bare relations are
quite simpler. Indeed, since $\eta _{\mathrm{B}}$ is dimensionless at $%
\varepsilon \neq 0$, while $\alpha _{\mathrm{B}}$ is dimensionful, $\eta _{%
\mathrm{B}}$ is just a constant, so 
\begin{equation}
\eta _{\mathrm{B}}(\alpha _{\mathrm{B}},\xi ,\varepsilon )=\zeta .
\label{costa}
\end{equation}
Re-written in terms of renormalized couplings, formula (\ref{costa}) gives 
\begin{equation}
\eta +\alpha \Delta _{\eta }(\alpha ,\eta ,\varepsilon )=\eta
+\sum_{k=1}^{\infty }\alpha ^{k}\widetilde{P}_{k+1}(\eta ,\varepsilon
)=\zeta ,  \label{costa2}
\end{equation}
where $\widetilde{P}_{k+1}(\eta ,\varepsilon )$ is a polynomial in $\eta $
of degree $k+1$ and a divergent function of $\varepsilon $. Formula (\ref
{costa2}) is an algebraic equation for $\eta (\alpha )$. The solution
coincides with the solution of (\ref{zimmesp}) once the constants $\xi $ and 
$\zeta $ are appropriately related to each other. Taking the limit $\alpha
\rightarrow 0$ of (\ref{costa2}), we obtain the relation between $\zeta $
and $\xi $, which reads 
\begin{equation}
\zeta (\varepsilon ,\xi )=\lim_{\alpha \rightarrow 0}\eta (\alpha ,\xi
,\varepsilon ),\qquad \varepsilon \neq 0.  \label{im}
\end{equation}
Using (\ref{im}), equation (\ref{costa2}) can be solved for $\eta (\alpha
,\xi ,\varepsilon )$ as an expansion in powers of $\alpha $. Thus, the
solution of the reduction equation is analytic in $\alpha $ at $\varepsilon
\neq 0$.

\bigskip

\textbf{Reduction of the renormalization constants}. Once $\eta $ is written
in terms of $\alpha $ and the arbitrary constant $\xi $, the divergences are
removed with a single renormalization constant. To show this fact, it is
convenient to work in an regularization framework (for example the cut-off
method or the Pauli-Villars regularization) where the bare couplings $\alpha
_{\mathrm{B}}$ and $\eta _{\mathrm{B}}$ coincide with the renormalized
couplings $\alpha _{\Lambda }$ and $\eta _{\Lambda }$ at the cut-off scale $%
\Lambda $. In such a framework, using the minimal subtraction scheme, the RG
consistency conditions are unaffected by the regularization, namely they
have the same form as (\ref{zimmesp}) at $\varepsilon =0$. Denote the
renormalization constants in this framework with $\widehat{Z}_{\alpha
}(\alpha ,\eta ,\ln \Lambda /\mu )$ and $\widehat{\Delta }_{\eta }(\alpha
,\eta ,\ln \Lambda /\mu )$. Then, because of the RG consistency conditions,
the relation $\eta =\eta (\alpha )$ holds at every energy scale, in
particular $\Lambda $ and the renormalization point $\mu $. We have 
\begin{equation}
\eta _{\mathrm{B}}=\eta (\alpha )+\alpha \widehat{\Delta }_{\eta }(\alpha
,\eta (\alpha ),\ln \Lambda /\mu )=\eta _{\Lambda }=\eta (\alpha _{\Lambda
})=\eta \left( \alpha \widehat{Z}_{\alpha }(\alpha ,\eta (\alpha ),\ln
\Lambda /\mu )\right) ,  \label{inside}
\end{equation}
and so the renormalization constant of $\eta $ is not independent, but
uniquely related to the one of $\alpha $: 
\begin{equation}
\widehat{\Delta }_{\eta }(\alpha ,\eta (\alpha ),\ln \Lambda /\mu )=\frac{1}{%
\alpha }\left[ \eta \left( \alpha \widehat{Z}_{\alpha }(\alpha ,\eta (\alpha
),\ln \Lambda /\mu )\right) -\eta (\alpha )\right] .  \label{undim}
\end{equation}
Formula (\ref{inside}) ensures that it is sufficient to renormalize the
coupling $\alpha $ \textit{inside }the function $\eta (\alpha )$ to remove
the divergences associated with $\eta $. In the dimensional-regularization
framework a relation between $\Delta _{\eta }$ and $Z_{\alpha }$ analogous
to (\ref{undim}) can be derived relating $\ln \Lambda /\mu $ to $%
1/\varepsilon $ and $\mu ^{\varepsilon }$.

The number of renormalization constants is reduced, because $\xi $ is finite
from the point of view of renormalization ($\xi _{\mathrm{B}}=\xi $, $Z_{\xi
}=1$). Instead, the number of independent couplings is not truly reduced,
because the constant $\xi $ is still arbitrary. Restrictions have to be
imposed on the solution $\eta (\alpha ,\xi ,\varepsilon )$ to have an
effective reduction. Because $\xi $ is finite, therefore a pure number, it
is meaningful to investigate criteria that fix the value of $\xi $
unambiguously.

\bigskip

\textbf{Leading-log approximation}. Before studying the general solution, it
is instructive to work out the solution in the leading-log approximation.
The one-loop beta functions have the form 
\begin{equation}
\beta _{\alpha }=\beta _{1}\alpha ^{2},\qquad \beta _{\eta }=\alpha \left(
a+b\eta +c\eta ^{2}\right) ,  \label{betaso}
\end{equation}
where $\beta _{1}$, $a$, $b$ and $c$ are numerical factors. The leading-log
solution of equation (\ref{zimmesp}) reads 
\begin{equation}
\eta _{\pm }(\alpha ,\xi ,\varepsilon )=-\frac{1}{2c}\left[ b\mp s\frac{%
1+\xi ~(-\varepsilon /Z_{\alpha })^{\pm s/\beta _{1}}}{1-\xi ~(-\varepsilon
/Z_{\alpha })^{\pm s/\beta _{1}}}\right] ,  \label{renoz}
\end{equation}
where 
\begin{equation}
Z_{\alpha }=\frac{1}{1-\alpha \beta _{1}/\varepsilon }  \label{za1}
\end{equation}
is the renormalization constant of $\alpha $ and $s$ is the square root of $%
b^{2}-4ac$. Here, $\eta _{\pm }$ are just two different ways to write the
same solution. If $b^{2}-4ac\geq 0$ we take $s$ to be the positive square
root.

For later use, it is convenient to invert the solution (\ref{renoz}) and
write $\xi $ as a function of $\alpha $ and $\eta $: 
\begin{equation}
\xi =(-\varepsilon /Z_{\alpha _{1}})^{\mp s/\beta _{1}}z_{\pm },\qquad \text{%
where }z_{\pm }=\frac{b\mp s+2c\eta _{\pm }}{b\pm s+2c\eta _{\pm }}.
\label{c1}
\end{equation}

The bare relations are (\ref{costa}). To write them explicitly, we calculate
the renormalization constants $\Delta _{\eta }(\alpha ,\eta ,\varepsilon )$
of $\eta $, using the procedure of the appendix. Basically, define 
\[
\widetilde{\Delta }_{\eta }(\alpha ,\xi ,\varepsilon )=\Delta _{\eta }\left(
\alpha ,\eta (\alpha ,\xi ,\varepsilon ),\varepsilon \right) , 
\]
then integrate the equation 
\[
\frac{\mathrm{d}\left( \alpha \widetilde{\Delta }_{\eta }\right) }{\mathrm{d}%
\alpha }=-\frac{\beta _{\eta }\left( \eta (\alpha ,\xi ,\varepsilon ),\alpha
\right) }{\beta _{\alpha }\left( \eta (\alpha ,\xi ,\varepsilon ),\alpha
\right) -\varepsilon \alpha } 
\]
along the RG\ flow, with the initial condition $\widetilde{\Delta }_{\eta
}(0,\xi ,\varepsilon )<\infty $. Finally insert (\ref{c1}) inside $%
\widetilde{\Delta }_{\eta }(\alpha ,\xi ,\varepsilon )$ to eliminate $\xi $.
The result is 
\begin{equation}
\Delta _{\eta }=-\frac{2}{\alpha }\frac{\left( a+b\eta +c\eta ^{2}\right)
\left( 1-Z_{\alpha }^{\pm s/\beta _{1}}\right) }{b\pm s+2c\eta -(b\mp
s+2c\eta )Z_{\alpha }^{\pm s/\beta _{1}}}.  \label{misuse}
\end{equation}
Using (\ref{abbare}), the constant $\zeta $ of (\ref{costa}) reads 
\begin{equation}
\zeta _{\pm }=-\frac{1}{2c}\left( b\mp s\frac{1+\xi (-\varepsilon )^{\pm
s/\beta _{1}}}{1-\xi (-\varepsilon )^{\pm s/\beta _{1}}}\right) .
\label{use}
\end{equation}
Note the similarities between this formula and (\ref{renoz}). It is useful
to invert (\ref{use}) and write also 
\begin{equation}
\xi =(-\varepsilon )^{\mp s/\beta _{1}}z_{\pm \mathrm{B}},\qquad z_{\pm 
\mathrm{B}}\equiv \frac{b\mp s+2c\eta _{\pm \mathrm{B}}}{b\pm s+2c\eta _{\pm 
\mathrm{B}}}.  \label{c1inv}
\end{equation}

Observe that if 
\begin{equation}
\pm \frac{s}{\beta _{1}}-1\notin \mathbb{N},  \label{ing}
\end{equation}
the relation (\ref{use}) the function $\zeta (\xi ,\varepsilon )$ is not
analytic, unless $\xi =0$ or $\xi =\infty $. Then the unique reductions that
are analytic both at the renormalized and bare levels are 
\[
\eta _{\pm }=-\frac{b\mp s}{2c}. 
\]
They are meaningful only if $s$ is real.

\bigskip

\textbf{General solution}. Now we study the function $\eta (\alpha ,\xi
,\varepsilon )$ beyond the leading-log approximation. We first prove that
the most general solution $\eta (\alpha ,\xi ,\varepsilon )$ of (\ref
{zimmesp}) is analytic in $\varepsilon $ at $\alpha \neq 0$. Insert the
expansion 
\begin{equation}
\eta (\alpha ,\xi ,\varepsilon )=\sum_{i=0}^{\infty }\varepsilon ^{i}\eta
_{i}(\alpha ,\xi )  \label{anxi}
\end{equation}
into (\ref{zimmesp}) and work out the equations for the $\eta _{i}$'s. The
equation for $\eta _{0}$ is just Zimmermann's equation 
\begin{equation}
a+b\eta _{0}+c\eta _{0}^{2}+\sum_{k=2}^{\infty }\alpha ^{k-1}Q_{k+1}(\eta
_{0})=\alpha \eta _{0}^{\prime }\left( \beta _{1}+\sum_{k=2}^{\infty }\alpha
^{k-1}P_{k}(\eta _{0})\right) .  \label{zero}
\end{equation}
Instead, $\eta _{i}$, $i>0$, obey the linear equations 
\begin{equation}
\alpha \overline{\beta }\eta _{i}^{\prime }=\eta _{i}\overline{\gamma }%
+\delta _{i}\left( \alpha ,\eta ,\eta ^{\prime }\right) ,  \label{i}
\end{equation}
where 
\[
\overline{\gamma }=b+2c\eta _{0}+\sum_{k=2}^{\infty }\alpha
^{k-1}Q_{k+1}^{\prime }(\eta _{0})-\eta _{0}^{\prime }\sum_{k=2}^{\infty
}\alpha ^{k}P_{k}^{\prime }(\eta _{0}),\qquad \overline{\beta }=\beta
_{1}+\sum_{k=2}^{\infty }\alpha ^{k-1}P_{k}(\eta _{0}), 
\]
and $\delta _{i}\left( \alpha ,\eta ,\eta ^{\prime }\right) $ is analytic in 
$\alpha $, polynomial in $\eta _{j},\eta _{j}^{\prime }$ with $j<i$ and does
not depend on $\eta _{k}$, $\eta _{k}^{\prime }$ with $k\geq i$.

Next we prove that if 
\begin{equation}
\pm \frac{s}{\beta _{1}}\notin \mathbb{N},  \label{ing2}
\end{equation}
then there exists a unique solution $\overline{\eta }(\alpha ,\varepsilon )$
that is analytic in $\alpha $ and $\varepsilon $. Indeed, (\ref{zero}) has
an analytic solution if (\ref{ing}) and \textit{a fortiori} (\ref{ing2})
hold. By induction in $i$, it is immediate to see that (\ref{i}) admit
unique solutions $\eta _{i}(\alpha )$ that are analytic in $\alpha $, again
if (\ref{ing2}) hold. Observe that (\ref{ing2}) is just slightly more
restrictive than (\ref{ing}), because it excludes also $s=0$.

Finally, we study the most general, non-analytic, solution. The solution of (%
\ref{zero}) has expansions \cite{zimme} 
\begin{equation}
\eta _{0}(\alpha ,\xi )=\eta _{\pm }(\alpha ,\xi ,0)=\sum_{k=0}^{\infty }%
\overline{c}_{\pm k}\alpha ^{k}+\sum_{n=1}^{\infty }\sum_{m=0}^{\infty }%
\overline{d}_{\pm mn}\xi ^{n}\alpha ^{m\pm ns/\beta _{1}},  \label{expa}
\end{equation}
where 
\[
\overline{c}_{\pm 0}=-\frac{b\mp s}{2c},\qquad \overline{d}_{\pm 01}=1, 
\]
and the other coefficients $\overline{c}_{\pm k}$, $\overline{d}_{\pm mn}$
are unambiguous calculable numbers. The solutions of (\ref{i}) can be worked
out inductively in $i$. Assume that the $\eta _{j}$'s, $j<i$, are known.
Then 
\[
\eta _{i}(\alpha ,\xi )=\int_{\overline{\alpha }_{i}}^{\alpha }\mathrm{d}%
\alpha ^{\prime }\frac{\delta _{i}\left( \alpha ^{\prime },\xi \right) \tau
(\alpha ,\alpha ^{\prime },\xi )}{\alpha ^{\prime }\overline{\beta }(\alpha
^{\prime },\xi )},\qquad \tau (\alpha ,\alpha ^{\prime },\xi )=\exp \left(
\int_{\alpha ^{\prime }}^{\alpha }\frac{\overline{\gamma }(\alpha ^{\prime
\prime },\xi )\mathrm{d}\alpha ^{\prime \prime }}{\alpha ^{\prime \prime }%
\overline{\beta }(\alpha ^{\prime \prime },\xi )}\right) , 
\]
where $\overline{\alpha }_{i}$ is a redundant arbitrary constant, that can
be fixed to any non-zero value. Observe that $\eta (\alpha ,\xi ,\varepsilon
)$ is analytic both in $\varepsilon $ and $\xi $ at $\alpha \neq 0$, an
important property that will be used later.

\bigskip

\textbf{Comparison of reductions}. The comparison between bare and
renormalized reduction relations can be studied generalizing formula (\ref
{expa}) and using tricks inspired by the $\varepsilon $-expansion. Let 
\begin{equation}
\alpha _{*}(\varepsilon )=\frac{\varepsilon }{\beta _{1}}+\mathcal{O}%
(\varepsilon ^{2}),\qquad \eta _{\pm *}(\varepsilon )=-\frac{b\mp s}{2c}+%
\mathcal{O}(\varepsilon ),  \label{starbeha}
\end{equation}
denote the non-trivial RG fixed point at $\varepsilon \neq 0$, namely the
solution of 
\[
\frac{1}{\alpha }\widehat{\beta }_{\alpha }(\alpha ,\eta ,\varepsilon
)=0,\qquad \frac{1}{\alpha }\beta _{\eta }(\alpha ,\eta ,\varepsilon )=0. 
\]
Using (\ref{abbe}) it is immediate to prove that the solutions are analytic
in $\varepsilon $ and have the behaviors (\ref{starbeha}), if $\beta
_{1}\neq 0$ and $s\neq 0$, which we assume. Define the new variables 
\[
u=\alpha -\alpha _{*}(\varepsilon ),\qquad v=\eta -\eta _{\pm *}(\varepsilon
) 
\]
and write expansions 
\[
\frac{\widehat{\beta }_{\alpha }}{\alpha }\equiv f(u,v)=f_{1}u+f_{2}v+%
\mathcal{O}(u^{2},uv,v^{2}),\qquad \frac{\beta _{\eta }}{\alpha }\equiv
g(u,v)=g_{1}u+g_{2}v+\mathcal{O}(u^{2},uv,v^{2}), 
\]
where $f_{1}=\beta _{1}+\mathcal{O}(\varepsilon )$, $g_{2}=\pm s+\mathcal{O}%
(\varepsilon )$, $f_{2}=\mathcal{O}(\varepsilon ^{2})$, $g_{1}=\mathcal{O}%
(1) $.

The reduction of couplings is expressed by a function $v(u)$ that satisfies 
\begin{equation}
f(u,v(u))\frac{\mathrm{d}v(u)}{\mathrm{d}u}=g(u,v(u)).  \label{rede}
\end{equation}
\newline
Consider also the equation 
\begin{equation}
f(k(v),v)=\frac{\mathrm{d}k(v)}{\mathrm{d}v}g(k(v),v).  \label{rede2}
\end{equation}
It is simple to see that if 
\begin{equation}
\pm \frac{\beta _{1}}{s}-1\notin \mathbb{N}  \label{banal}
\end{equation}
then (\ref{rede2}) admits an analytic solution 
\begin{equation}
k(v)=v\sum_{k=0}^{\infty }b_{k}v^{k},  \label{anal}
\end{equation}
such that the coefficients $b_{k}$ are analytic functions of $\varepsilon $
and $b_{k}=\mathcal{O}(\varepsilon ^{2})$. This $\varepsilon $-behavior can
be proved observing that formulas (\ref{abbe}) imply 
\[
\left. \frac{\partial ^{m+n}f}{\partial u^{m}\partial v^{n}}\right| _{u=v=0}=%
\mathcal{O}(\varepsilon ^{n-m}),\qquad \left. \frac{\partial ^{m+n+2}g}{%
\partial u^{m}\partial v^{n+2}}\right| _{u=v=0}=\mathcal{O}(\varepsilon
^{n-m}), 
\]
if $m\leq n$. Use the solution (\ref{anal}) to define $u^{\prime }=u-k(v)$.
Then equations (\ref{rede}) can be rewritten in the form 
\begin{equation}
\widetilde{f}(u^{\prime },v(u^{\prime }))\ u^{\prime }\frac{\mathrm{d}%
v(u^{\prime })}{\mathrm{d}u^{\prime }}=\widetilde{g}(u^{\prime },v(u^{\prime
})),  \label{redep}
\end{equation}
where $\widetilde{f}$ and $\widetilde{g}$ are analytic functions of $%
u^{\prime }$ and $v$. Precisely $\widetilde{g}(u^{\prime },v)=g(u,v)$ and 
\[
\widetilde{f}(u^{\prime },v)=\frac{1}{u^{\prime }}\left[ f(u^{\prime
}+k(v),v)-\frac{\mathrm{d}k}{\mathrm{d}v}g(u^{\prime }+k(v),v)\right] =%
\widetilde{f}_{1}+\mathcal{O}(u^{\prime },v), 
\]
and $\widetilde{f}_{1}=\beta _{1}+\mathcal{O}(\varepsilon )$. Finally, it is
easy to see that the solution of (\ref{redep}) has expansions 
\begin{equation}
v_{\pm }(u^{\prime },\xi )=\sum_{k=1}^{\infty }c_{\pm k}^{\prime }(u^{\prime
})^{k}+\sum_{n=1}^{\infty }\sum_{m=0}^{\infty }d_{\pm mn}^{\prime }\xi
^{n}(u^{\prime })^{m+nQ_{\pm }},\qquad Q_{\pm }=\pm \frac{s}{\beta _{1}}+%
\mathcal{O}(\varepsilon ),  \label{compa}
\end{equation}
where the coefficients $c_{\pm k}^{\prime }$, $d_{\pm mn}^{\prime }$ are
unambiguous calculable functions of $\varepsilon $ ($d_{\pm 01}^{\prime }$
being set to 1), and $\xi $ is the arbitrary constant. If $s/\beta _{1}>0$
the meaningful expansions are $v_{+}(u^{\prime },\xi )$ and $v_{-}(u^{\prime
},0)$, if $s/\beta _{1}<0$ they are $v_{+}(u^{\prime },0)$ and $%
v_{-}(u^{\prime },\xi )$.

The invertibility conditions (\ref{banal}), necessary to write the solution $%
k(v)$ and (\ref{compa}), are not included in (\ref{ing2}). Typically, $s$ is
irrational or complex, so it is not difficult to fulfill both (\ref{ing2})
and (\ref{banal}).

Using (\ref{compa}) and the variable changes performed so far, it is
possible to express the solution (\ref{compa}) in the $\alpha $-$\eta $
parametrization, and write $\eta (\alpha ,\xi ,\varepsilon )$. It is then
easy to show again that $\eta (\alpha ,\xi ,\varepsilon )$ is analytic both
in $\varepsilon $ and $\xi $ at $\alpha \neq 0$.

\bigskip

\textbf{Relation between the renormalized and bare arbitrary constants }$\xi 
$\textbf{\ and }$\zeta $. Inserting (\ref{compa}) into (\ref{im}) the $%
\alpha \rightarrow 0$ limit gives an implicit equation for $\zeta _{\pm }$,
that can be solved recursively in powers of $\varepsilon $ and $\xi
\varepsilon ^{Q_{\pm }}$. Observe that 
\[
\lim_{\alpha \rightarrow 0}u^{\prime }=-\alpha _{*}(\varepsilon )-k\left(
\zeta _{\pm }-\eta _{\pm *}(\varepsilon )\right) =\varepsilon \left( -\frac{1%
}{\beta _{1}}+\mathcal{O}(\varepsilon ,\varepsilon \zeta _{\pm },...)\right)
. 
\]
The relation $\zeta _{\pm }(\varepsilon ,\xi )$ is non-analytic in $%
\varepsilon $ at $\xi \neq 0$, so the double-analyticity requirement
correctly implies $\xi =0$ and selects the reduction uniquely.

In the leading-log approximation the known results can be immediately
recovered.

\bigskip

\textbf{Relation between the two criteria for the reduction}. The limit (\ref
{im}) not only explains the similarities between $\eta (\alpha ,\xi
,\varepsilon )$ and $\zeta (\varepsilon ,\xi )$, exhibited by (\ref{renoz})
and (\ref{use}) in the leading-log approximation, but also establishes a
connection between the two criteria for the reduction, namely analyticity in 
$\alpha $ and $\varepsilon $ at the origin, and analyticity in $\varepsilon $
at the bare and renormalized levels. The interpolation between the two
requirements is encoded in the function $u^{\prime }=\alpha -\varepsilon
/\beta _{1}+\cdots $. Raised to non-integer powers, $u^{\prime }$ originates
both the non-integer powers of $\alpha $ in the $\varepsilon \rightarrow 0$
limit and the non-integer powers of $\varepsilon $ in the $\alpha
\rightarrow 0$ limit. The ``dual'' roles played by $\alpha $ and $%
\varepsilon $ are not surprising, if we recall that in the perturbative
regime 
\[
\ln \Lambda \sim \frac{1}{\varepsilon },\qquad \alpha \sim \frac{1}{\ln
\Lambda },
\]
where $\Lambda $ denotes a cut-off, so $\alpha \sim \varepsilon $.

\section{Infinite reduction}

\setcounter{equation}{0}

In this section we study the infinite reduction for non-renormalizable
theories at the regularized level, using the dimensional-regularization
technique. The minimal subtraction scheme is used for the unreduced theory.
The subtraction scheme of the reduced theory is the one induced by the
reduction itself. Unless otherwise specified, the words ``relevant'',
``marginal'' and ``irrelevant'' refer to the Gaussian fixed point, so they
are equivalent to ``super-renormalizable'', ``strictly renormalizable'' and
``non-renormalizable'', respectively. In the study of deformations of
interacting conformal field theories, our construction allows also to
characterize the deformation as marginal, relevant or irrelevant at the
interacting fixed point.

For definiteness, we work in four dimensions. The generalization to odd and
other even dimensions is direct and left to the reader. Let $\mathcal{R}$
denote the power-counting renormalizable subsector of the theory. For the
moment we assume that $\mathcal{R}$ contains a single marginal coupling $%
\alpha =g^{2}$, where $\alpha $ multiplies the four-leg marginal vertices
and $g$ multiplies the three-leg marginal vertices. The power-counting
renormalizable sector $\mathcal{R}$ needs to be fully interacting, because
the infinite reduction does not work when the marginal sector is free or
only partially interacting. We assume also that $\mathcal{R}$ does not
contain relevant couplings. This assumption ensures that the beta functions
of the irrelevant sector depend polynomially on the irrelevant couplings, so
there exists a simple recursive procedure to solve the reduction equations.
When $\mathcal{R}$ is fully interacting, relevant parameters can be added 
\textit{after} the construction of the irrelevant deformation and studied
perturbatively in the reduction equations. In practice, interactions have to
be turned on in the following order: first the marginal interactions, then
the irrelevant interactions, finally the relevant interactions.

Let $\mathcal{O}_{n}$ denote a basis of local, ``essential'', scalar,
symmetric, canonically irrelevant operators of $\mathcal{R}$. Essential
operators are defined as the equivalence classes of operators that differ by
total derivatives, terms proportional to the field equations and BRST-exact
terms \cite{wein}. Total derivatives are trivial in perturbation theory. The
terms proportional to the field equations can be renormalized away by means
of field redefinitions. Finally, the BRST-exact sector does not affect the
physical quantities. ``Symmetric'' means that the integrated operators have
to be invariant under the non-anomalous symmetries of the theory.

The irrelevant terms can be ordered according to their ``level''. If, at $%
\varepsilon =0$, the operator $\mathcal{O}_{n}$ has canonical dimensionality 
$d_{n}$ in units of mass, then the level $n$ of $\mathcal{O}_{n}$ is the
difference $d_{n}-D$, $D$ being the physical spacetime dimension. In
general, each level contains finitely many operators $\mathcal{O}_{n}^{I}$,
which can mix under renormalization.

It is convenient to write the classical lagrangian in the form 
\begin{equation}
\mathcal{L}_{\mathrm{cl}}[\varphi ]=\mathcal{L}_{\mathcal{R}}[\varphi
,g]+\sum_{n>0}\sum_{I}g^{2p_{n}^{I}}\lambda _{n}^{I}\mathcal{O}%
_{n}^{I}(\varphi ),  \label{nonra}
\end{equation}
where $\varphi $ generically denotes the fields of the theory, $\mathcal{L}_{%
\mathcal{R}}[\varphi ,g]$ is the lagrangian of the renormalizable subsector
and $p_{n}^{I}=N_{n}^{I}/2-1$, where $N_{n}^{I}$ is the number of legs of $%
\mathcal{O}_{n}^{I}$. The bare couplings are 
\begin{equation}
\alpha _{\mathrm{B}}=g_{\mathrm{B}}^{2}=\alpha \mu ^{\varepsilon }Z_{\alpha
}(\alpha ,\varepsilon ),\qquad \lambda _{n\mathrm{B}}^{I}=\lambda
_{n}^{J}Z_{n}^{IJ}(\alpha ,\lambda ,\varepsilon ),  \label{aba}
\end{equation}
and the bare lagrangian reads 
\[
\mathcal{L}_{\mathrm{cl~B}}[\varphi _{\mathrm{B}}]=\mathcal{L}_{\mathcal{R~}%
\mathrm{B}}[\varphi _{\mathrm{B}},g_{\mathrm{B}}]+\sum_{n>0}\sum_{I}\lambda
_{n\mathrm{B}}^{I}g_{\mathrm{B}}^{2p_{n}^{I}}\mathcal{O}_{n}^{I}(\varphi _{%
\mathrm{B}}). 
\]

Call \textit{dimensionality-defect }$p$ of a quantity the difference between
its dimensionality at $\varepsilon \neq 0$ and its dimensionality at $%
\varepsilon =0$, divided by $\varepsilon $. For example, the
dimensionality-defect of $\alpha _{\mathrm{B}}$ is $1$ and the one of $g_{%
\mathrm{B}}$ is 1/2. Assume that the kinetic terms of the fields are
conventionally multiplied by unity. Then a \textit{minimal }coupling $\chi $
is the coefficient of a vertex. In symbolic notation, the vertex reads 
\[
\chi [\partial ^{q}]\phi ^{n_{s}}\psi ^{n_{f}}A^{n_{v}}G^{n_{g}}, 
\]
where $[\partial ^{q}]$ stands for $q$ variously distributed derivatives, $%
n_{s}$, $n_{f}$, $n_{v}$ and $n_{g}$ are the numbers of scalar, fermion,
vector and graviton legs, respectively. The dimensionality-defect of $\chi _{%
\mathrm{B}}$ is 
\begin{equation}
p_{\chi }=\frac{N}{2}-1,\qquad N=n_{s}+n_{f}+n_{v}+n_{g},  \label{piennet}
\end{equation}
where $N$ is the total number of legs. Thus, $p_{\chi }$ is greater than
zero whenever $N>2$. One-leg terms are associated with scalar vacuum
expectation values and the cosmological term. The two-leg terms include mass
terms, kinetic terms and contributions of the cosmological term. In very
general situations, including gravity \cite{renprop}, the higher-derivative
kinetic terms can be converted into vertices, mass terms and the
cosmological term, using the field equations. As mentioned above, in the
construction of irrelevant deformations by means of the infinite reduction,
the relevant parameters are initially turned off (they can be turned on
perturbatively at a secondary stage). Therefore, for the purposes of the
infinite reduction, we can assume that there are no independent quadratic
terms besides the free kinetic ones and that the vertices have three legs or
more, which ensures that every essential coupling has $p>0$. The factors $%
g^{2p_{n}^{I}}$ of formula (\ref{nonra}) have been introduced so that the
dimensionality-defects of the non-minimal irrelevant couplings $\lambda
_{n\ell \mathrm{B}}^{I}$ are zero. Observe that, by definition, also the
marginal vertices contained in $\mathcal{R}$ are multiplied by $g$-powers
equal to the number of legs minus two.

It is useful to define a parity transformation $U$, that sends $g$ into $-g$
and every field $\varphi $ into $-\varphi $. Clearly, $\mathcal{R}$ is $U$%
-invariant. Then it is evident that (\ref{nonra}) is $U$-invariant and each $%
\lambda _{n}^{I}$ is $U$-invariant. The parametrization (\ref{nonra}) is
convenient because it allows us to work with $U$-even quantities.

Using the minimal subtraction scheme, the RG equations read 
\begin{equation}
\frac{\mathrm{d}\alpha }{\mathrm{d}\ln \mu }=\widehat{\beta }_{\alpha
}(\alpha ,\varepsilon )=\beta _{\alpha }\left( \alpha \right) -\varepsilon
\alpha ,\qquad \frac{\mathrm{d}\lambda _{n}^{I}}{\mathrm{d}\ln \mu }=\beta
_{n}^{I}\left( \alpha ,\lambda \right) .  \label{abe}
\end{equation}
The beta function of the irrelevant couplings $\lambda _{n}$ have the form 
\begin{equation}
\beta _{n}^{I}(\lambda ,\alpha )=\gamma _{n}^{IJ}(\alpha )~\lambda
_{n}^{J}+\alpha \delta _{n}^{I}(\alpha ,\lambda ),  \label{betagel}
\end{equation}
where $\delta _{n}^{I}(\alpha ,\lambda )$ depends polynomially, at least
quadratically, on the irrelevant couplings $\lambda _{m}$ with $m<n$ and
does not depend on the irrelevant couplings with $m\geq n$. Instead, $\gamma
_{n}^{IJ}(\alpha )$ is the matrix of anomalous dimensions of the operators $%
g^{2p_{n}^{I}}\mathcal{O}_{n}^{I}(\varphi )$ (defined up to total
derivatives and terms proportional to the field equations), calculated in
the undeformed theory $\mathcal{R}$. Both $\delta _{n}^{I}(\alpha ,\lambda )$
and $\gamma _{n}^{IJ}(\alpha )$ are analytic in $\alpha $ and $\gamma
_{n}^{IJ}(\alpha )$ is of order $\alpha $.

The structure (\ref{betagel}) follows from dimensional analysis and simple
diagrammatics. In perturbation theory only non-negative powers of the
couplings can appear and by assumption the theory does not contain
parameters with positive dimensionalities in units of mass. Then, matching
the dimensionalities of the left- and right-hand sides of (\ref{betagel}),
the $\lambda $-dependence of the right-hand side of (\ref{betagel}) follows.
As far as the $\alpha $-dependence is concerned, let $G$ be a diagram
contributing to the renormalization of the vertex $\mathcal{O}%
_{n}^{J}(\varphi )$, with $E=N_{n}^{J}$ external legs, $I$ internal legs and 
$V$ vertices. The $g$-powers carried by the (marginal and irrelevant)
vertices due to the non-minimal parametrization (\ref{nonra}) are equal to 
\begin{equation}
\sum_{\text{vertices}}\left( \#\text{legs}-2\right) =E+2I-2V,  \label{legs}
\end{equation}
recalling that the total number of legs attached to the vertices is $E+2I$.
Since $I-V=L-1\geq 0$, where $L$ is the number of loops, the total $g$-power
carried by the diagram is $N_{n}^{J}-2+2L=2p_{n}^{J}+2L$. Thus 
\begin{equation}
\mu ^{-p_{n}^{J}\varepsilon }g_{\mathrm{B}}^{2p_{n}^{J}}\lambda _{n\mathrm{B}%
}^{J}=g^{2p_{n}^{J}}\lambda _{n}^{J}+\sum_{L\geq 1}g^{2p_{n}^{J}}\alpha
^{L}d_{n,L}^{J}(\varepsilon ,\lambda ),  \label{uy}
\end{equation}
where $d_{n,L}^{J}(\varepsilon ,\lambda )$ are divergent coefficients that
do not depend on the $\lambda _{k}$'s with $k>n$. Since $Z_{g}=Z_{\alpha
}^{1/2}=1+\sum_{L\geq 1}\alpha ^{L}c_{L}(\varepsilon )$, (\ref{uy}) gives
immediately 
\begin{equation}
\lambda _{n\mathrm{B}}^{J}=\lambda _{n}^{J}+\sum_{L\geq 1}\alpha
^{L}d_{n,L}^{\prime \ J}(\varepsilon ,\lambda ),\qquad \lambda
_{n}^{J}=\lambda _{n\mathrm{B}}^{J}+\sum_{L\geq 1}\mu ^{-\varepsilon
L}\alpha _{\mathrm{B}}^{L}d_{n,L}^{\prime \prime \ J}(\varepsilon ,\lambda _{%
\mathrm{B}}),  \label{uo}
\end{equation}
for other divergent coefficients $d_{n,L}^{\prime \ J}$, $d_{n,L}^{\prime
\prime \ J}$ that do not depend on the $\lambda _{k}$'s with $k>n$. Thus the
beta function of $\lambda _{n}^{J}$ has the $\alpha $-dependence specified
in (\ref{betagel}).

Another way to derive (\ref{betagel}) is as follows. Define rescaled fields $%
\varphi =\varphi ^{\prime }/g$. Then each primed field is $U$-even and (\ref
{nonra}) becomes 
\begin{equation}
\mathcal{L}_{\mathrm{cl}}^{\prime }[\varphi ^{\prime }]=\frac{1}{\alpha }%
\left[ \mathcal{L}_{\mathcal{R}}^{\prime }[\varphi ^{\prime
}]+\sum_{n}\sum_{J}\lambda _{n}^{J}\mathcal{O}_{n}^{J}(\varphi ^{\prime
})\right] .  \label{appare}
\end{equation}
In this parametrization, every propagator carries a factor $\alpha $ and
every vertex carries a factor $1/\alpha $. So, a diagram with $I$ internal
legs, $L$ loops and $V$ vertices carries a factor $\alpha ^{I-V}=\alpha
^{L}/\alpha $, that gives a counterterm contributing to $\lambda _{n\mathrm{B%
}}^{J}/\alpha _{\mathrm{B}}$. Thus (\ref{uo}) follow immediately and the
beta function of $\lambda _{n}^{J}$ is a sum of contributions proportional
to $\alpha ^{L}$, $L\geq 1$, in agreement with (\ref{betagel}).

Observe that the structures (\ref{nonra}) and (\ref{appare}) are compatible
both with gauge invariance and the use of field equations.

The $\alpha $-beta function and the anomalous dimensions of $\mathcal{R}$
have expansions 
\begin{equation}
\beta _{\alpha }=\alpha ^{2}\beta _{\alpha }^{(1)}+\mathcal{O}(\alpha
^{3}),\qquad \gamma _{n}(\alpha )=\alpha \gamma _{n}^{(1)}+\mathcal{O}%
(\alpha ^{2}),  \label{tz2}
\end{equation}
etc. and we assume $\beta _{\alpha }^{(1)}\neq 0$. For the moment it is
convenient to ignore the indices labelling operators of the same level,
which amounts to discard the renormalization mixing.

Now we study the infinite reduction. As usual, the reduced deformation is
made of a head and a queue. The head is the irrelevant term of lowest
dimensionality, whose level is denoted with $\ell $. The queue is made of
terms of levels $n\ell $, with $n$ integer. The head is multiplied by the
irrelevant coupling $\lambda _{\ell }$, while the terms of the queue are
multiplied by functions of $\lambda _{\ell }$ and the marginal couplings $%
\alpha $ of $\mathcal{R}$. Write 
\begin{equation}
\mathcal{L}_{\mathrm{cl}}[\varphi ]=\mathcal{L}_{\mathcal{R}}[\varphi
,g]+g^{2p_{\ell }}\lambda _{\ell }\mathcal{O}_{\ell }(\varphi
)+\sum_{n>1}g^{2p_{n\ell }}\lambda _{n\ell }(\alpha ,\lambda _{\ell
},\varepsilon )~\mathcal{O}_{n\ell }(\varphi ).  \label{th}
\end{equation}

\bigskip

\textbf{Renormalized reduction relations.} On dimensional grounds, the
renormalized reduction relations have the form 
\begin{equation}
\lambda _{n\ell }(\alpha ,\lambda _{\ell },\varepsilon )=\lambda _{\ell
}^{n}f_{n}(\alpha ,\varepsilon ),\qquad n>1.  \label{eq}
\end{equation}
The lowest level of the deformation has $\delta _{\ell }=0$, so 
\begin{equation}
\beta _{\ell }(\lambda ,\alpha ,\varepsilon )=\lambda _{\ell }~\gamma _{\ell
}(\alpha ).  \label{eq1}
\end{equation}

Assume inductively that the functions $f_{m}(\alpha ,\xi ,\varepsilon )$ are
known for $m<n$ and that they depend on certain constants $\xi _{m}$, $m<n$.
Then, since $\delta _{n}(\alpha ,\lambda )$ depends only on the irrelevant
couplings $\lambda _{m\ell }$ with $m<n$, it is possible to write 
\begin{equation}
\delta _{n\ell }(\alpha ,\lambda ,\varepsilon )=\overline{\delta }%
_{n}(\alpha ,\xi ,\varepsilon )~\lambda _{\ell }^{n},  \label{ura}
\end{equation}
where $\overline{\delta }_{n}(\alpha ,\xi ,\varepsilon )$ are known
functions that depend on $\xi _{k}$ with $k<n$. Differentiating (\ref{eq})
and using (\ref{betagel}) and (\ref{eq1}) we obtain the equation 
\begin{equation}
f_{n}^{\prime }(\alpha ,\varepsilon )~\widehat{\beta }_{\alpha
}=f_{n}(\alpha ,\varepsilon )~\left( \gamma _{n\ell }(\alpha )-n\gamma
_{\ell }(\alpha )\right) +\alpha \overline{\delta }_{n}(\alpha ,\xi
,\varepsilon ).  \label{diffeq}
\end{equation}
These are the RG\ consistency conditions for the functions $f_{n}(\alpha
,\varepsilon )$. They are first order differential equations, so the
solutions contain arbitrary constants $\xi _{n}$, one for every $n$. Since
the beta function of $\lambda _{n\ell }$ depends only on the $\lambda
_{k\ell }$'s with $k\leq n$, the function $f_{n}(\alpha ,\xi ,\varepsilon )$
depends only the constants $\xi _{k}$ with $k\leq n$. The solution of (\ref
{diffeq}) can be split into a sum of two terms, 
\begin{equation}
f_{n}(\alpha ,\xi ,\varepsilon )=\overline{f}_{n}(\alpha ,\xi ,\varepsilon
)+\xi _{n}\overline{s}_{n}(\alpha ,\varepsilon ),  \label{split}
\end{equation}
where $\overline{s}_{n}(\alpha ,\varepsilon )$ solves the homogeneous
equation, $\xi _{n}$ is the arbitrary integration constant belonging to $%
f_{n}$ and $\overline{f}_{n}(\alpha ,\xi ,\varepsilon )$ is a particular
solution, that depends on the constants $\xi _{k}$ with $k<n$.

If $\overline{\delta }_{n}\neq 0$ the solution can be written also as 
\begin{equation}
f_{n}(\alpha ,\xi ,\varepsilon )=\int_{\xi _{n}^{\prime }}^{\alpha }\mathrm{d%
}\alpha ^{\prime }\frac{\alpha ^{\prime }\overline{\delta }_{n}(\alpha
^{\prime },\xi ,\varepsilon )~s_{n}(\alpha ,\alpha ^{\prime },\varepsilon )}{%
\widehat{\beta }_{\alpha }(\alpha ^{\prime },\varepsilon )},  \label{sola}
\end{equation}
where 
\begin{equation}
s_{n}(\alpha ,\alpha ^{\prime },\varepsilon )=\exp \left( \int_{\alpha
^{\prime }}^{\alpha }\mathrm{d}\alpha ^{\prime \prime }\frac{\gamma _{n\ell
}(\alpha ^{\prime \prime })-n\gamma _{\ell }(\alpha ^{\prime \prime })}{%
\widehat{\beta }_{\alpha }(\alpha ^{\prime \prime },\varepsilon )}\right)
\label{sn}
\end{equation}
and the $\xi _{n}^{\prime }$'s are constants suitably related to the $\xi
_{n}$'s.

If $\beta _{\alpha }\neq 0$ the solutions are analytic in $\varepsilon $.
Indeed, assume by induction that $f_{m}(\alpha ,\xi ,\varepsilon )$, $m<n$,
are analytic in $\varepsilon $. Then $\overline{\delta }_{n}(\alpha ,\xi
,\varepsilon )$ is analytic in $\varepsilon $ and (\ref{sola})-(\ref{sn})
show that also $f_{n}(\alpha ,\xi ,\varepsilon )$ is analytic in $%
\varepsilon $.

At $\varepsilon \neq 0$ the solutions are also analytic in $\alpha $.
Indeed, for $\alpha $ small $\widehat{\beta }_{\alpha }(\alpha ^{\prime
},\varepsilon )\sim -\alpha ^{\prime }\varepsilon $. Assume by induction
that $f_{m}(\alpha ,\xi ,\varepsilon )$, $m<n$, are analytic in $\alpha $.
Then $\overline{\delta }_{n}(\alpha ,\xi ,\varepsilon )$ is analytic in $%
\alpha $ and (\ref{sola})-(\ref{sn}) show that also $f_{n}(\alpha ,\xi
,\varepsilon )$ is analytic in $\alpha $.

From the point of view of renormalization, the constants $\xi $ are finite
arbitrary parameters ($Z_{\xi }=1$) and the divergences of (\ref{th}) are
removed by means of renormalization constants for $\alpha $ and $\lambda
_{\ell }$, plus field redefinitions, with no independent renormalization
constants for the couplings $\lambda _{n\ell }$, $n>1$, of the queue. This
fact can be proved with an argument analogous to the one leading to (\ref
{undim}) (see also \cite{infredfin}).

Now we prove that if certain invertibility conditions are fulfilled, then
there exists a unique solution $\overline{f}_{n}(\alpha ,\varepsilon )$ that
is analytic both in $\alpha $ and $\varepsilon $.

\bigskip

\textbf{Doubly analytic solution}. If the invertibility conditions 
\begin{equation}
\tau _{k}\equiv \frac{\gamma _{k\ell }^{(1)}-k\gamma _{\ell }^{(1)}}{\beta
_{\alpha }^{(1)}}\notin \mathbb{N},\qquad k>1  \label{inveps}
\end{equation}
hold, then there exists a unique solution $\overline{f}_{k}(\alpha
,\varepsilon )$ that is analytic in $\alpha $ and $\varepsilon $. Use the $%
\varepsilon $-analyticity of $f_{n}(\alpha ,\varepsilon )$ at $\alpha \neq 0$
to write the expansion 
\begin{equation}
\overline{f}_{k}(\alpha ,\varepsilon )=\sum_{i=0}^{\infty }\varepsilon ^{i}%
\overline{f}_{i,k}(\alpha ).  \label{form}
\end{equation}
Assume inductively that (\ref{form}) are analytic both in $\alpha $ and $%
\varepsilon $ for $k<n$. Then the functions $\overline{f}_{i,k}(\alpha )$, $%
k<n$, are analytic in $\alpha $ and we can write 
\begin{equation}
\overline{\delta }_{n}(\alpha ,\varepsilon )=\sum_{i=0}^{\infty }\varepsilon
^{i}\overline{\delta }_{i,n}(\alpha ),  \label{deltaform}
\end{equation}
where $\overline{\delta }_{i,n}(\alpha )$ are analytic in $\alpha $. Insert (%
\ref{form}) and (\ref{deltaform}) into (\ref{diffeq}) and write the
reduction equations as 
\begin{equation}
\beta _{\alpha }\frac{\mathrm{d}\overline{f}_{i,n}(\alpha )}{\mathrm{d}%
\alpha }=\overline{f}_{i,n}(\alpha )~\left( \gamma _{n\ell }(\alpha
)-n\gamma _{\ell }(\alpha )\right) +\alpha \frac{\mathrm{d}\overline{f}%
_{i-1,n}(\alpha )}{\mathrm{d}\alpha }+\alpha \overline{\delta }_{i,n}(\alpha
),  \label{arreda}
\end{equation}
$i=0,1,\ldots $, with $\overline{f}_{-1,n}(\alpha )=0$. The solution of (\ref
{arreda}) can be worked out recursively in $i$ and, for given $i$, in power
series of $\alpha $. It is then immediate to see that if the invertibility
conditions (\ref{inveps}) hold, there exist unique analytic solutions $%
\overline{f}_{i,n}(\alpha )$.

\bigskip

The arbitrary constant $\xi _{n}$ multiplies the function $\overline{s}%
_{n}(\alpha ,\varepsilon )$, which is not analytic in both $\alpha $ and $%
\varepsilon $, when the invertibility conditions are fulfilled. Indeed, $%
\overline{s}_{n}(\alpha ,\varepsilon )$ is proportional to $s_{n}(\alpha ,%
\overline{\alpha },\varepsilon )$, with $\overline{\alpha }>0$. Expanding $%
s_{n}(\alpha ,\overline{\alpha },\varepsilon )$ in powers of $\varepsilon $
it is immediate to see that the terms of the expansion are not analytic in $%
\alpha $ around $\alpha \sim 0$. Instead, at $\varepsilon \neq 0$ $%
s_{n}(\alpha ,\overline{\alpha },\varepsilon )$ is obviously analytic in $%
\alpha $, since it is just a product of renormalization constants of the
undeformed theory $\mathcal{R}$.

\bigskip

\textbf{Violations of the invertibility conditions}. When some invertibility
conditions are violated, namely $\tau _{\overline{n}}=\overline{r}\in %
\mathbb{N}$ for some $\overline{n}$, then it is necessary to introduce a new
independent coupling. Consider (\ref{arreda}). The solution $\overline{f}_{0,%
\overline{n}}(\alpha )$ can be worked out in power series of $\alpha $ up to
the order $\alpha ^{\overline{r}-1}$, while the coefficient of $\alpha ^{%
\overline{r}}$ is ill-defined: the problem is avoided introducing a new
independent coupling $\lambda _{\overline{n}\ell }^{(0)}$ at order $\alpha ^{%
\overline{r}}$. Similarly, the solutions $\overline{f}_{i,\overline{n}%
}(\alpha )$, $0<i<\overline{r}$, can be worked out up to the orders $\alpha
^{\overline{r}-i-1}$, but they require new couplings $\lambda _{\overline{n}%
\ell }^{(i)}$ from the orders $\alpha ^{\overline{r}-i}$ on. However, only $%
\lambda _{\overline{n}\ell }^{(0)}$ is a new physical coupling, because the $%
\lambda _{\overline{n}\ell }^{(i)}$'s with $i>0$ belong to the evanescent
sector, so they do not affect the physical quantities. These properties
ensure that the violations of the invertibility conditions are less harmful
than they appear at first sight: certainly they cause the introduction of
new parameters, possibly infinitely many, but in general $\tau _{n}$ grows
with $n$ and the physical new couplings appear at higher and higher orders,
thus permitting low-order predictions with a relatively small number of
independent couplings \cite{infredfin}.

It is convenient to introduce $\overline{r}+1$ new parameters $\lambda _{%
\overline{n}\ell }^{(i)}$, $i=0,1,\ldots \overline{r}$, and write 
\begin{equation}
\lambda _{\overline{n}\ell }=f_{\overline{n}}(\alpha ,\varepsilon )\lambda
_{\ell }^{\overline{n}}+\sum_{i=0}^{\overline{r}}\alpha ^{\overline{r}%
-i}\varepsilon ^{i}\lambda _{\overline{n}\ell }^{(i)},\qquad \widehat{\beta }%
_{\overline{n}_{\ell }}^{(i)}=\gamma _{\overline{n}\ell }^{(i)}\lambda _{%
\overline{n}\ell }^{(i)}+\varepsilon (\overline{r}-i)\lambda _{\overline{n}%
\ell }^{(i)}+\alpha \lambda _{\ell }^{\overline{n}}\delta _{\overline{n}\ell
}^{(i)}(\alpha ,\varepsilon ),  \label{viola}
\end{equation}
where $f_{\overline{n}}(\alpha ,\varepsilon )$ is determined up to the
orders $\alpha ^{\overline{r}-1-i}\varepsilon ^{i}$, $i=0,\ldots \overline{r}%
-1$, solving the reduction equations (\ref{arreda}), $\gamma _{\overline{n}%
\ell }^{(i)}=\gamma _{\overline{n}\ell }-(\overline{r}-i)\beta _{\alpha
}/\alpha =\mathcal{O}(\alpha )$ and $\delta _{\overline{n}\ell }^{(i)}$ are
analytic in $\alpha $. As anticipated above, the new physical parameter $%
\lambda _{\overline{n}\ell }^{(0)}$ is multiplied by $\alpha ^{\overline{r}}$%
. For $n>\overline{n}$ write 
\begin{equation}
\lambda _{n\ell }=\sum_{\{m\}}f_{n,\{m\}}(\alpha ,\varepsilon )~\lambda
_{\ell }^{\widehat{m}}\prod_{i=0}^{\overline{r}}\left( \alpha ^{\overline{r}%
-i}\varepsilon ^{i}\lambda _{\overline{n}\ell }^{(i)}\right) ^{m_{i}},
\label{viola2}
\end{equation}
where $\widehat{m},m_{i}$ are integers such that $\widehat{m}+\overline{n}%
\sum_{i=0}^{\overline{r}}m_{i}=n$. In (\ref{betagel}) $\delta _{n}(\alpha
,\lambda )$ can be decomposed as 
\[
\delta _{n}(\alpha ,\lambda )=\sum_{\{m\}}\delta _{n,\{m\}}(\alpha
,\varepsilon )~\lambda _{\ell }^{\widehat{m}}\prod_{i=0}^{\overline{r}%
}\left( \alpha ^{\overline{r}-i}\varepsilon ^{i}\lambda _{\overline{n}\ell
}^{(i)}\right) ^{m_{i}}. 
\]
Then (\ref{betagel}) give equations of the form 
\begin{equation}
\widehat{\beta }_{\alpha }f_{n,\{m\}}^{\prime }=\left( \gamma _{n\ell }-%
\widehat{m}\gamma _{\ell }-\sum_{j=0}^{\overline{r}}m_{j}\gamma _{\overline{n%
}\ell }\right) f_{n,\{m\}}+\alpha \widehat{\delta }_{n,\{m\}}(\alpha
,f,\varepsilon )  \label{fnm}
\end{equation}
where $\widehat{\delta }_{n,\{m\}}(\alpha ,f,\varepsilon )$ depends on the
functions $f_{k,\{m^{\prime }\}}$ with $k<n$ and $f_{n,\{m^{\prime }\}}$
with $\widehat{m}^{\prime }<\widehat{m}$. The invertibility conditions are
still (\ref{inveps}) for $n>\overline{n}$, because the one-loop coefficient
of the combination of anomalous dimensions written in the parenthesis of (%
\ref{fnm}) is equal to 
\[
\gamma _{n\ell }^{(1)}-n\gamma _{\ell }^{(1)}-\beta _{\alpha }^{(1)}%
\overline{r}\sum_{j=0}^{\overline{r}}m_{j}. 
\]
Then, it is possible to solve (\ref{fnm}) recursively in $\widehat{m}$ for
given $n$ and there exist unique solutions $f_{n,\{m\}}(\alpha ,\varepsilon
) $ that are analytic in $\alpha $ and $\varepsilon $.

We remark that the appearance of new parameters is guided by the
construction itself. Moreover, the invertibility conditions (\ref{inveps})
depend only on the one-loop $\mathcal{R}$ beta function and the one-loop
anomalous dimensions of the composite operators of $\mathcal{R}$. Such
quantities are calculable in the undeformed renormalizable subsector $%
\mathcal{R}$, before turning the irrelevant deformation on. Thus, it is
possible to count the parameters of the non-renormalizable interaction, or
say how dense they are, before constructing the non-renormalizable
interaction. This property emphasized once again the perturbative character
of the infinite reduction.

\bigskip

\textbf{Renormalization mixing. }Taking into account of the renormalization
mixing, the analysis generalizes in a simple way. Assume for the moment that
the head does not mix. Distinguish the couplings of the same level with
extra indices $I,J,\ldots $ Write $\lambda _{n\ell }^{I}(\alpha ,\lambda
_{\ell },\varepsilon )=\lambda _{\ell }^{n}f_{n}^{I}(\alpha ,\varepsilon )$, 
$n>1$, $\overline{f}_{n}^{I}(\alpha ,\varepsilon )=\sum_{i=0}^{\infty
}\varepsilon ^{i}\overline{f}_{i,n}^{I}(\alpha )$. Formula (\ref{diffeq})
becomes 
\begin{equation}
\beta _{\alpha }\frac{\mathrm{d}\overline{f}_{i,n}^{I}(\alpha )}{\mathrm{d}%
\alpha }=\left( \gamma _{n\ell }^{IJ}(\alpha )-n\delta ^{IJ}\gamma _{\ell
}(\alpha )\right) \overline{f}_{i,n}^{J}(\alpha )+\alpha \frac{\mathrm{d}%
\overline{f}_{i-1,n}^{I}(\alpha )}{\mathrm{d}\alpha }+\alpha \overline{%
\delta }_{i,n}^{I}(\alpha ),  \label{equamix}
\end{equation}
which admit unique analytic solutions $\overline{f}_{i,n}^{I}(\alpha )$ if
the matrices 
\begin{equation}
\frac{\gamma _{n\ell }^{(1)IJ}-n\delta ^{IJ}\gamma _{\ell }^{(1)}}{\beta
_{\alpha }^{(1)}},  \label{invmix}
\end{equation}
$n>1$, have no integer eigenvalue.

If the head itself has a non-trivial mixing with other operators, call $%
\gamma _{\ell }^{(1)}$ a real eigenvalue of $\gamma _{\ell }^{(1)IJ}$. We
assume for simplicity that the eigenvalue $\gamma _{\ell }^{(1)}$ has
multiplicity one. Perform a constant redefinition $\lambda _{\ell
}^{I}\rightarrow M^{IJ}\lambda _{\ell }^{J}$, $M^{IJ}$=constant, to put the
matrix $\gamma _{\ell }^{(1)IJ}$ into its canonical Jordan form with $\gamma
_{\ell }^{(1)NN}=\gamma _{\ell }^{(1)}$, $\gamma _{\ell }^{(1)N\overline{J}%
}=\gamma _{\ell }^{(1)\overline{I}N}=0$. Here the unoverlined indices $I$, $%
J $ range from $1$ to $N$ and the overlined indices $\overline{I}$, $%
\overline{J}$ range from 1 to $N-1$. Then take $\lambda _{\ell }\equiv
\lambda _{\ell }^{N}$ as independent coupling and reduce the other level-$%
\ell $ couplings as 
\[
\lambda _{\ell }^{\overline{I}}=f^{\overline{I}}(\alpha ,\varepsilon
)\lambda _{\ell }. 
\]
The level-$\ell $ beta functions $\beta _{\ell }^{I}=\gamma _{\ell
}^{IJ}\lambda _{\ell }^{J}$ give 
\begin{equation}
\beta _{\ell }=\beta _{\ell }^{N}=\left( \gamma _{\ell }^{NN}+\gamma _{\ell
}^{N\overline{I}}f^{\overline{I}}\right) \lambda _{\ell },\qquad \widehat{%
\beta }_{\alpha }\frac{\mathrm{d}f^{\overline{I}}}{\mathrm{d}\alpha }=\left(
\gamma _{\ell }^{\overline{I}\overline{J}}-\delta ^{\overline{I}\overline{J}%
}\gamma _{\ell }^{NN}\right) f^{\overline{J}}+\gamma _{\ell }^{\overline{I}%
N}-f^{\overline{I}}\gamma _{\ell }^{N\overline{J}}f^{\overline{J}}.
\label{ammix}
\end{equation}
The second equation depends quadratically on the $f^{\overline{I}}$'s, but
the quadratic term is multiplied by $\gamma _{\ell }^{N\overline{J}}$, which
is $\mathcal{O}(\alpha ^{2})$ by construction. Writing $f^{\overline{I}%
}(\alpha ,\varepsilon )=\sum_{i=0}^{\infty }\varepsilon ^{i}f_{i}^{\overline{%
I}}(\alpha )$ as usual, the equations for $f_{i}^{\overline{I}}(\alpha )$
can be solved recursively in $i$ and in power series of $\alpha $. The
doubly analytic solution exists and is unique if the matrices 
\[
\frac{\gamma _{\ell }^{(1)\overline{I}\overline{J}}-\delta ^{\overline{I}%
\overline{J}}\gamma _{\ell }^{(1)}}{\beta _{\alpha }^{(1)}} 
\]
have no integer eigenvalue. The functions $f^{\overline{I}}$ determine also
the beta function $\beta _{\ell }$. Observe that the anomalous dimension $%
\gamma _{\ell }\equiv \gamma _{\ell }^{NN}+\gamma _{\ell }^{N\overline{I}}f^{%
\overline{I}}$ of the coupling $\lambda _{\ell }$ is equal to $\alpha \gamma
_{\ell }^{(1)}+\mathcal{O}(\alpha ^{2})$. At higher levels the reduction
proceeds as usual and the invertibility conditions are still that the
matrices (\ref{invmix}) have no integer eigenvalue for $n>1$. The head of
the deformation is $\sum_{I}g^{2p_{\ell }^{I}}\lambda _{\ell }^{I}\mathcal{O}%
_{\ell }^{I}(\varphi )$.

If the eigenvalue $\gamma _{\ell }^{(1)}$ is complex it is necessary to
consider a two-head deformation involving also its complex conjugate \cite
{infredfin}.

\bigskip

\textbf{Perturbative meromorphy.} The reduced theory reads 
\begin{equation}
\mathcal{L}_{\mathrm{cl}}[\varphi ]=\mathcal{L}_{\mathcal{R}}[\varphi
,g]+\sum_{n=1}^{\infty }\sum_{I}g^{2p_{n\ell }^{I}}\lambda _{\ell }^{n}%
\overline{f}_{n}^{I}(\alpha ,\varepsilon )~\mathcal{O}_{n\ell }^{I}(\varphi
).  \label{redth}
\end{equation}
Since $p_{n\ell }^{I}>0$ every term of the irrelevant deformation is
parametrized in a non-minimal way and in the $g\rightarrow 0$ limit at fixed 
$\lambda _{\ell }$ the theory becomes free. In this parametrization, $%
\lambda _{\ell }=1/M_{P\mathrm{eff}}^{\ell }$ defines the \textit{effective
Planck mass }$M_{P\mathrm{eff}}$ and the perturbative expansion in powers of
the energy $E$ is meaningful for $E\ll M_{P\mathrm{eff}}$. On the other
hand, define the \textit{Planck mass} $M_{P}=g^{-2\overline{p}/\ell }\lambda
_{\ell }^{-1/\ell }$, in such a way that the irrelevant terms with
dimensionality-defect $\overline{p}$ are coupled in a minimal way. Then we
get 
\begin{equation}
\mathcal{L}[\varphi ]=\mathcal{L}_{\mathcal{R}}[\varphi
,g]+\sum_{n=1}^{\infty }\sum_{I}g^{2\widetilde{p}_{n\ell }^{I}}\overline{f}%
_{n}^{I}(\alpha ,\varepsilon )~M_{P}^{-n\ell }~\mathcal{O}_{n\ell
}^{I}(\varphi ),  \label{tyupin}
\end{equation}
where $\widetilde{p}_{n\ell }^{I}=p_{n\ell }^{I}-n\overline{p}$. Most of the
numbers $\widetilde{p}_{n\ell }^{I}$ are negative, so the $g\rightarrow 0$
limit at fixed $M_{P}$ is singular. Nevertheless, the singularity is bounded
by the order of the perturbative expansion and indeed can be reabsorbed into
the effective Planck mass. Because of this feature, the reduction is said to
be \textit{perturbatively meromorphic} \cite{infredfin}. Since the $g$%
-singularities of (\ref{tyupin}) can be reabsorbed only in a fully
non-minimal parametrization, there is no way to turn the marginal
interaction off, keeping the irrelevant interaction on, which is why the
renormalizable sector $\mathcal{R}$ needs to be fully interacting.

\section{Bare infinite reduction}

\setcounter{equation}{0}

In this section we study the infinite reduction at the bare level and relate
it to the renormalized infinite reduction studied in the previous section.
We want to derive an alternative criterion to select the infinite reduction.
In the previous section we showed that the infinite reduction follows from
double analyticity in $\alpha $ and $\varepsilon $ of the renormalized
reduction relations. Here we do not pay attention to the $\alpha $%
-dependence and show that $\overline{f}_{k}(\alpha ,\varepsilon )$ is also
the unique solution such that both the bare and renormalized reduction
relations are analytic in $\varepsilon $.

\bigskip

\textbf{Bare reduction relations.} At the level of bare couplings, the
reduction has a simpler form, namely 
\begin{equation}
\lambda _{n\ell \mathrm{B}}=\zeta _{n}\lambda _{\ell \mathrm{B}}^{n},
\label{abr}
\end{equation}
where the $\zeta _{n}$'s are constants. This expression is fixed matching
the naive dimensionalities and the dimensionality-defects and demanding
analyticity in $\lambda _{\ell \mathrm{B}}$ and $\alpha _{\mathrm{B}}$.
Since $\lambda _{n\ell \mathrm{B}}$ and $\lambda _{\ell \mathrm{B}}$ are
dimensionless at $\varepsilon \neq 0$ and $\alpha _{\mathrm{B}}$ is
dimensionful, the bare relations do not depend on $\alpha _{\mathrm{B}}$.

Now, rewrite (\ref{abr}) in terms of the renormalized couplings, using (\ref
{aba}): 
\begin{equation}
\lambda _{n\ell }Z_{n\ell }(\alpha ,\lambda ,\varepsilon )=\zeta _{n}\lambda
_{\ell }^{n}Z_{\ell }^{n}(\alpha ,\varepsilon ).  \label{redux2}
\end{equation}
These are algebraic equations relating $\lambda _{n\ell }$ with $\lambda
_{\ell }$ and $\alpha $. In particular, they can be used to work out another
expression of $f_{n}(\alpha ,\xi ,\varepsilon )$, that must be equivalent to
(\ref{split}) and (\ref{sola}) once $\zeta $ and $\xi $ are suitably related.

\bigskip

The main goal of this section is to prove that if the invertibility
conditions (\ref{inveps}) are fulfilled, there exists unique values of $\xi
_{n}$ and $\zeta _{n}$ such that both (\ref{split}) and (\ref{abr}) are
analytic in $\varepsilon $. In our notation these values are $\xi _{n}=0$
for every $n$, while $\zeta _{n}$ are equal to suitable analytic functions $%
\overline{\zeta }_{n}(\varepsilon )$ of $\varepsilon $.

\bigskip

\textbf{Equivalence of the bare and renormalized reduction equations.} First
we prove that the algebraic relations (\ref{redux2}) are equivalent to the
differential equations (\ref{diffeq}), once $\xi _{n}$ and $\zeta _{n}$ are
appropriately related to each other.

Coherently with (\ref{betagel}), it is convenient to decompose $\lambda
_{n\ell }Z_{n\ell }$ into a sum of two contributions: the terms that are
linear in $\lambda _{n\ell }$ plus the terms that are at least quadratic in
the irrelevant couplings with lower levels. Precisely, 
\begin{equation}
\lambda _{n\ell }Z_{n\ell }(\alpha ,\lambda ,\varepsilon )\equiv
z_{n}(\alpha ,\varepsilon )\left[ \lambda _{n\ell }+\alpha \Delta
_{n}(\alpha ,\lambda ,\varepsilon )\right] ,  \label{boccia}
\end{equation}
where $z_{n}(\alpha ,\varepsilon )=1+\mathcal{O}(\alpha )$ is the
renormalization constant of $\mathcal{O}_{n\ell }(\varphi )$, viewed as a
composite operator of the undeformed theory $\mathcal{R}$, while $\Delta
_{n}(\alpha ,\lambda ,\varepsilon )$ depends only on $\lambda _{k\ell }$
with $k<n$ and is analytic in $\alpha $. Indeed, we have proved in the
previous section that every counterterm that renormalizes the product $%
\alpha ^{p_{n\ell }}\lambda _{n\ell }$ carries a power $\alpha ^{p_{n\ell
}+L}$, where $L$ is the number of loops. Thus the counterterms that
renormalize $\lambda _{n\ell }$ are proportional to $\alpha ^{L}$, $L\geq 1$.

Inserting (\ref{eq}) into (\ref{boccia}) and (\ref{redux2}), and defining $%
\Delta _{n}(\alpha ,\lambda ,\varepsilon )=\lambda _{\ell }^{n}\overline{%
\Delta }_{n}(\alpha ,f,\varepsilon )$, where $\overline{\Delta }_{n}(\alpha
,f,\varepsilon )$ depends only on $f_{k}$ with $k<n$, we get 
\begin{equation}
f_{n}(\alpha ,\varepsilon )=-\alpha \overline{\Delta }_{n}(\alpha
,f,\varepsilon )+\zeta _{n}z_{n}^{-1}(\alpha ,\varepsilon )Z_{\ell
}^{n}(\alpha ,\varepsilon ).  \label{redux}
\end{equation}
This formula iteratively gives the functions $f_{n}$, $n>1$, in terms of $%
\alpha ,\zeta ,\varepsilon $, and $f_{k}$ with $k<n$. By direct
differentiation, it is immediate to verify that the functions $f_{n}(\alpha
,\varepsilon )$ of (\ref{redux}) do satisfy also (\ref{diffeq}). Thus, the
solutions $f_{n}(\alpha ,\xi ,\varepsilon )$ of (\ref{diffeq}) must coincide
with (\ref{redux}), once the constants $\xi $ are written in terms of $\zeta 
$, or vice versa. Observe that formula (\ref{redux}) shows again that the
solution is analytic in $\alpha $ at $\varepsilon \neq 0$.

A quicker way to show that\ the bare relations (\ref{abr}) integrate the
renormalized reduction equations (\ref{diffeq}) is as follows. Use (\ref{abr}%
) to write 
\[
\zeta _{n}=\frac{\lambda _{n\ell \mathrm{B}}}{\lambda _{\ell \mathrm{B}}^{n}}%
. 
\]
The right-hand side of this formula, rewritten in terms of renormalized
couplings, is clearly independent of the renormalization point $\mu $.
Therefore $\zeta _{n}$ is an integral of motion of the RG flow. Since the
renormalized reduction equations (\ref{diffeq}) are just ratios of RG
equations (see (\ref{zimmesp})), $\lambda _{n\ell }(\lambda _{\ell },\alpha
) $ determined from (\ref{redux2}) solves (\ref{diffeq}).

\bigskip

\textbf{Relation between the bare and renormalized constants }$\zeta $, $\xi 
$\textbf{\ and uniqueness of the doubly analytic reduction. }The particular
solution (\ref{form}) is analytic in $\alpha $ and $\varepsilon $. We now
prove that, without paying attention to the $\alpha $-dependence, the
solution (\ref{form}) is identified uniquely also by the requirement that
both the bare and renormalized reduction relations be analytic in $%
\varepsilon $. For this purpose it is useful to work out the structure of
the function $\zeta _{n}(\xi ,\varepsilon )$.

First observe that

\begin{equation}
\lim_{\alpha \rightarrow 0}\overline{\Delta }_{n}(\alpha ,f(\alpha ,\xi
,\varepsilon ),\varepsilon )<\infty \text{,\qquad at }\varepsilon \neq 0.
\label{state}
\end{equation}
Indeed, every $f(\alpha ,\xi ,\varepsilon )$ is analytic in $\alpha $ at $%
\varepsilon \neq 0$ and $\overline{\Delta }_{n}(\alpha ,f,\varepsilon )$ is
analytic in $\alpha $.

Applying (\ref{state}) to (\ref{redux}) and recalling that $z_{n}$ and $%
Z_{\ell }$ tend to one when $\alpha $ tends to zero, we obtain a useful
formula to compute the relation between the bare constants $\zeta _{n}$ and
the renormalized constants $\xi _{n}$, namely 
\begin{equation}
\zeta _{n}(\xi ,\varepsilon )=\lim_{\alpha \rightarrow 0}f_{n}(\alpha ,\xi
,\varepsilon )\text{,\qquad }\varepsilon \neq 0.  \label{limit}
\end{equation}

Now, $\zeta _{n}$ is linear in $\xi _{n}$, so it is convenient to split it
into the sum of two contributions, 
\begin{equation}
\zeta _{n}(\xi ,\varepsilon )=\overline{\zeta }_{n}(\xi ,\varepsilon )+\xi
_{n}\widehat{\zeta }_{n}(\varepsilon ),  \label{split2}
\end{equation}
to be studied separately, such that $\overline{\zeta }_{n}(\varepsilon ,\xi
) $ depends on the $\xi _{k}$'s with $k<n$.

\bigskip

\textbf{Pole cancellations in }(\ref{redux2})\textbf{.} The algebraic
equations (\ref{redux2}) contain poles. However, (\ref{redux}) solves (\ref
{diffeq}), and we know that (\ref{diffeq}) admits a solution that is
analytic in $\varepsilon $ at $\alpha \neq 0$. Therefore, the poles of (\ref
{redux2}) have to mutually cancel out, once $\zeta _{n}$ is replaced by the
appropriate function $\zeta _{n}(\xi ,\varepsilon )$ (\ref{split2}). The
mechanism of pole cancellation provides an alternative method to derive the
results obtained in the previous section, including the invertibility
conditions (\ref{inveps}). We describe the cancellation in the case that is
most interesting for our purposes, that is to say the doubly analytic
solution $\overline{f}_{n}(\alpha ,\varepsilon )$, which corresponds to $\xi
=0$, i.e. $\zeta _{n}=\zeta _{n}(0,\varepsilon )=\overline{\zeta }%
_{n}(0,\varepsilon )\equiv \overline{\zeta }_{n}(\varepsilon )$. We prove
also that the function $\overline{\zeta }_{n}(\varepsilon )$ is analytic in $%
\varepsilon $.

First observe that the divergences of $\overline{\Delta }_{n}$, $z_{n}$ and $%
Z_{\ell }$ are consistent with the renormalization-group. This is true also
of $\overline{\Delta }_{n}(\alpha ,\overline{f}(\alpha ,\varepsilon
),\varepsilon )$, when the functions $f_{k}$, $k<n$, are inductively
replaced by the solutions $\overline{f}_{k}(\alpha ,\varepsilon )$ of the
reduction relations, because the reduction equations are just ratios of RG\
equations. Consequently, the double and higher poles of (\ref{redux}) are
unambiguously related to the simple poles and it is sufficient to check that
the simple poles of (\ref{redux}) cancel out to prove complete cancellation.

We proceed inductively. Assume that $\zeta _{k}$, for $k<n$, are equal to
analytic functions $\overline{\zeta }_{k}(\varepsilon )$ of $\varepsilon $
such that $\overline{f}_{k}(\alpha ,\varepsilon )\equiv -\alpha \overline{%
\Delta }_{k}(\alpha ,\overline{f},\varepsilon )+\overline{\zeta }%
_{k}(\varepsilon )z_{k}^{-1}(\alpha ,\varepsilon )Z_{\ell }^{k}(\alpha
,\varepsilon )$ are analytic in $\alpha $ and $\varepsilon $. Write $%
\overline{\zeta }_{n}(\varepsilon )=\sum_{k=0}^{\infty }$ $\overline{\zeta }%
_{n,k}\varepsilon ^{k}$. In (\ref{redux}) the coefficient $\overline{\zeta }%
_{n,k}$ is multiplied by a sum of objects of the form 
\begin{equation}
\varepsilon ^{k}\left( \frac{\alpha }{\varepsilon }\right) ^{m}\alpha ^{r},
\label{lead}
\end{equation}
with $m,r\geq 0$. The simple pole is 
\begin{equation}
\frac{1}{\varepsilon }\alpha ^{1+k+r}.  \label{buc}
\end{equation}
Since $\overline{f}_{k}(\alpha ,\varepsilon )$, $k<n$, are analytic in $%
\alpha $ and $\varepsilon $ by the inductive hypothesis, the simple pole of $%
\overline{\Delta }_{n}$ is an analytic function of $\alpha $. In total, the
simple poles of (\ref{redux}) have the form 
\begin{equation}
\frac{\alpha }{\varepsilon }\left( \sum_{s\geq 0}a_{s}\alpha
^{s}+\sum_{k,r\geq 0}\overline{\zeta }_{n,k}c_{k,r}\alpha ^{k+r}\right) ,
\label{pare}
\end{equation}
where $a_{s}$ and $c_{k,r}$ are known numerical factors. If the coefficients
of $\overline{\zeta }_{n,j}\alpha ^{j}$ inside the parenthesis are nonzero
it is possible to uniquely determine $\overline{\zeta }_{n,j}$ iteratively
in $j$ from the cancellation of the pole. The coefficient of $\overline{%
\zeta }_{n,j}\alpha ^{j}$ depends only on the leading-log contributions to
the wave-function renormalization constants, given by the standard formulas 
\cite{collins} 
\begin{equation}
Z_{\alpha }=\left( 1-\frac{\beta _{\alpha }^{(1)}\alpha }{\varepsilon }%
\right) ^{-1},\qquad Z_{\ell }=Z_{\alpha }^{\gamma _{\ell }^{(1)}/\beta
_{\alpha }^{(1)}},\qquad z_{n}=Z_{\alpha }^{\gamma _{n\ell }^{(1)}/\beta
_{\alpha }^{(1)}}.  \label{llog}
\end{equation}
Inside the parenthesis of (\ref{pare}) $\overline{\zeta }_{n,j}\alpha ^{j}$
is multiplied by the coefficient 
\[
\frac{\left( -\beta _{\alpha }^{(1)}\right) ^{j+1}}{(j+1)!}%
\prod_{i=0}^{j}\left( \frac{\gamma _{n\ell }^{(1)}-n\gamma _{\ell }^{(1)}}{%
\beta _{\alpha }^{(1)}}-i\right) , 
\]
which must not vanish. Thus we recover the invertibility conditions (\ref
{inveps}) from pole cancellation. We conclude that the doubly analytic
solution is determined by an $\varepsilon $-analytic constant $\zeta _{n}=%
\overline{\zeta }_{n}(\varepsilon )$.

\bigskip

Now we study the $\xi $-dependent terms of (\ref{split2}) and show that if
the invertibility conditions (\ref{inveps}) hold, when some $\xi $ is
nonzero either the renormalized or the bare relations are not analytic in $%
\varepsilon $. It is sufficient to prove the statement for $\xi _{n}\neq 0$
and $\xi _{k}=0$ for $k<n$.

Comparing (\ref{sn}) with (\ref{redux}), and using (\ref{split2}), we obtain 
\[
\xi _{n}\overline{s}_{n}(\alpha ,\varepsilon )=\xi _{n}\widehat{\zeta }%
_{n}(\varepsilon )z_{n}^{-1}(\alpha ,\varepsilon )Z_{\ell }^{n}(\alpha
,\varepsilon )=\xi _{n}\widehat{\zeta }_{n}(\varepsilon )s_{n}(\alpha
,0,\varepsilon ), 
\]
where $s_{n}(\alpha ,\alpha ^{\prime },\varepsilon )$ is given in (\ref{sn}%
). Now, in general $s_{n}(\alpha ,0,\varepsilon )$ is not analytic in $%
\varepsilon $, but $s_{n}(\alpha ,\overline{\alpha },\varepsilon )$
certainly is, if $\overline{\alpha }\neq 0$ and $\beta _{\alpha }(\alpha
)\neq 0$. Formula (\ref{split}) can be written as 
\begin{equation}
f_{n}(\alpha ,\xi ,\varepsilon )=\overline{f}_{n}(\alpha ,\varepsilon )+\xi
_{n}\widehat{\zeta }_{n}(\varepsilon )s_{n}(\overline{\alpha },0,\varepsilon
)s_{n}(\alpha ,\overline{\alpha },\varepsilon ).  \label{redux0}
\end{equation}

Assume that the invertibility conditions (\ref{inveps}) are fulfilled. If $%
\widehat{\zeta }_{n}(\varepsilon )$ is an analytic function of $\varepsilon $%
, then the bare relations (\ref{abr}) are analytic, but the renormalized
ones (\ref{redux0}) are not, because $s_{n}(\overline{\alpha },0,\varepsilon
)$ is not analytic in $\varepsilon $. The renormalized relations (\ref
{redux0}) become analytic choosing $\widehat{\zeta }_{n}(\varepsilon )=%
\widetilde{\zeta }_{n}(\varepsilon )s_{n}(0,\overline{\alpha },\varepsilon )$%
, with $\widetilde{\zeta }_{n}(\varepsilon )$ analytic. Then, however, the
bare relations (\ref{abr}) are not analytic.

Concluding, when the invertibility conditions are fulfilled, the bare and
renormalized reduction relations are both analytic in $\varepsilon $ if and
only if $\xi _{n}=0$, $\zeta _{n}=\overline{\zeta }_{n}(\varepsilon )$. This
condition uniquely determines the reduction.

\bigskip

\textbf{Renormalization mixing. }Taking into account of the renormalization
mixing in the bare reduction, formulas (\ref{abr}) and (\ref{boccia})
generalize to 
\begin{equation}
\lambda _{n\ell \mathrm{B}}^{I}=Z_{n\ell }^{IJ}(\alpha ,\lambda ,\varepsilon
)\lambda _{n\ell }^{J}\equiv \sum_{J}z_{n}^{IJ}(\alpha ,\varepsilon )\left[
\lambda _{n\ell }^{J}+\alpha \Delta _{n}^{J}(\alpha ,\lambda ,\varepsilon
)\right] =\zeta _{n}^{I}\lambda _{\ell \mathrm{B}}^{n},  \label{amix}
\end{equation}
where the coupling $\lambda _{\ell }$ is determined by the same equation (%
\ref{amix}) for $n=1$, which we write as 
\begin{equation}
\lambda _{\ell \mathrm{B}}^{I}=\sum_{J}z^{IJ}\lambda _{\ell }^{J}=\zeta
^{I}\lambda _{\ell \mathrm{B}}.  \label{uno}
\end{equation}
Assume that the coefficient-matrix $\gamma _{\ell }^{(1)IJ}$ of the one-loop
anomalous dimensions is arranged into its Jordan canonical form. With the
same notational conventions as in the previous section, write $I=(\overline{I%
},N)$, $\overline{I}=1,\ldots N-1$, $\gamma _{\ell }^{(1)\overline{I}%
N}=\gamma _{\ell }^{(1)N\overline{I}}=0$ and assume that the eigenvalue $%
\gamma _{\ell }^{(1)NN}\equiv \gamma _{\ell }^{(1)}$ is real and has
multiplicity one. Choose $\zeta ^{N}=1$ and define $\lambda _{\ell \mathrm{B}%
}^{N}=\lambda _{\ell \mathrm{B}}=Z_{\ell }\lambda _{\ell }$, $\lambda _{\ell
}^{N}=\lambda _{\ell }$, $\lambda _{\ell }^{\overline{I}}=f^{\overline{I}%
}\lambda _{\ell }$. Then (\ref{uno}) gives 
\begin{equation}
Z_{\ell }=z^{NN}+z^{N\overline{I}}f^{\overline{I}},\qquad \qquad f^{%
\overline{I}}=(z^{-1})^{\overline{I}\overline{J}}\left( z^{NN}\zeta ^{%
\overline{J}}+\zeta ^{\overline{J}}z^{N\overline{K}}f^{\overline{K}}-z^{%
\overline{J}N}\right) ,  \label{fs}
\end{equation}
where $z^{NN}=1+\mathcal{O}(\alpha )$, $z^{N\overline{I}}=\mathcal{O}(\alpha
^{2})$, $z^{\overline{I}N}=\mathcal{O}(\alpha ^{2})$, $z^{\overline{I}%
\overline{J}}=\delta ^{\overline{I}\overline{J}}+\mathcal{O}(\alpha )$.

Writing $\lambda _{n\ell }^{I}(\alpha ,\lambda _{\ell },\varepsilon
)=\lambda _{\ell }^{n}f_{n}^{I}(\alpha ,\varepsilon )$, for $n>1$, (\ref
{redux}) is replaced by 
\begin{equation}
f_{n}^{I}(\alpha ,\xi ,\varepsilon )=-\alpha \overline{\Delta }%
_{n}^{I}(\alpha ,f,\varepsilon )+\sum_{J}\left( z_{n}^{-1}(\alpha
,\varepsilon )\right) ^{IJ}\zeta _{n}^{J}Z_{\ell }^{n}(\alpha ,\varepsilon ).
\label{fo}
\end{equation}
The doubly analytic solution can be worked out repeating the analysis of
pole cancellation. The invertibility conditions are again that the matrices (%
\ref{invmix}) have no integer eigenvalue for $n>1$ and no positive integer
eigenvalue for $n=1$. The second equation of (\ref{fs}) is solved in powers
of $\alpha $, which determines the functions $f^{\overline{I}}$. Then the
first formula of (\ref{fs}) gives $Z_{\ell }$, which is inserted in (\ref{fo}%
). Finally, the equations (\ref{fo}) are solved for $f_{n}^{I}$, $n>1$.

Repeating the arguments leading to (\ref{limit}), we find also 
\[
\zeta _{n}^{I}(\xi ,\varepsilon )=\lim_{\alpha \rightarrow
0}f_{n}^{I}(\alpha ,\xi _{n},\varepsilon ). 
\]
Again, when the invertibility conditions are fulfilled the bare and
renormalized reduction relations cannot be both analytic in $\varepsilon $
at $\xi _{n}\neq 0$ .

\bigskip

\textbf{Reduced subtraction scheme.} Observe that the reduction of couplings
reduces also the scheme freedom. Indeed, a scheme change is a
reparametrization in the space of couplings. Before the reduction, every
coupling is independent and can be reparametrized independently. After the
reduction, instead, only $\alpha $ and $\lambda _{\ell }$ are independent.
Under an $\alpha ,\lambda _{\ell }$-reparametrization the queue of the
irrelevant deformation is reparametrized consistently. That is why, even if
the minimal subtraction scheme is used for the unreduced theory, various
evanescent terms appear after the reduction, due to the $\varepsilon $%
-dependence of the functions $\overline{f}_{n}(\alpha ,\varepsilon )$.
Explicit examples are given in the next sections.

\bigskip

Concluding, when the invertibility conditions are fulfilled, the reduced
theory is described by the classical lagrangian (\ref{redth}), whose
structure is preserved by renormalization. The divergences are subtracted
away with field redefinitions and a finite number of independent
renormalization constants:\ those belonging to the renormalizable sector $%
\mathcal{L}_{\mathcal{R}}[\varphi ,g]$ plus a renormalization constant for
the coupling $\lambda _{\ell }$ that multiplies the head of the deformation.
The bare lagrangian reads 
\begin{equation}
\mathcal{L}_{\mathrm{B}}[\varphi _{\mathrm{B}}]=\mathcal{L}_{\mathcal{R}%
}[\varphi _{\mathrm{B}},g_{\mathrm{B}}]+\sum_{n=1}^{\infty }\sum_{I}g_{%
\mathrm{B}}^{2p_{n\ell }^{I}}\overline{\zeta }_{n}^{I}(\varepsilon )\lambda
_{\ell \mathrm{B}}^{n}~\mathcal{O}_{n\ell }^{I}(\varphi _{\mathrm{B}}).
\label{defar2}
\end{equation}
Formulas (\ref{redth})\ and (\ref{defar2}) define \textit{the} fundamental
interaction having head $\sum_{I}g^{2p_{\ell }^{I}}\lambda _{\ell }^{I}%
\mathcal{O}_{\ell }^{I}(\varphi )$.

\bigskip

We have also seen that the invertibility conditions for the existence of the
reduction to all orders are captured just by the leading-log approximation.
Remarkably, the leading-log approximation can be solved exactly and provides
a good illustration of the properties found so far. This calculation is done
in the section 6.

\section{\textbf{Physical invertibility conditions in the absence of
three-leg marginal vertices}}

\setcounter{equation}{0}

When there are no three-leg vertices, e.g. $\mathcal{R}$ is the theory $%
\varphi ^{4}$ in four dimensions (but similar arguments apply if $\mathcal{R}
$ is the theory $\varphi ^{6}$ in three dimensions), it is possible to
refine the previous analysis and recover the results of \cite
{renscal,infredfin}. It is useful to introduce the numbers $\widetilde{p}%
_{n}\equiv p_{n\ell }-np_{\ell }$, which are always integers, by $U$-parity.
Indeed, if the head $\mathcal{O}_{\ell }(\varphi )$ is $U$-even, then the
terms of the queue are $U$-even, so the $\widetilde{p}_{n}$'s are integers.
If the head is $U$-odd, then the terms of the queue of levels $n\ell $ with $%
n$ odd are $U$-odd, while those with $n$ even are $U$-even. In both cases, $%
\widetilde{p}_{n}$ is integer.

Define the integer 
\begin{equation}
q_{k}\equiv \max (-k+1-\widetilde{p}_{k},0).  \label{qk}
\end{equation}
We prove that the general form of $\overline{\zeta }_{n}(\varepsilon )$ is 
\begin{equation}
\overline{\zeta }_{n}(\varepsilon )=\varepsilon ^{q_{n}}\sum_{k=0}^{\infty }%
\overline{\zeta }_{n,k}\varepsilon ^{k},  \label{anx}
\end{equation}
while the general form of $\overline{f}_{k}(\alpha ,\varepsilon )$ is 
\begin{equation}
\overline{f}_{k}(\alpha ,\varepsilon )=\alpha ^{q_{k}}\sum_{j=0}^{q_{k}}%
\overline{f}_{j,k}(\alpha ,\varepsilon )\left( \frac{\varepsilon }{\alpha }%
\right) ^{j},  \label{form2}
\end{equation}
where $\overline{f}_{j,k}(\alpha ,\varepsilon )$ are analytic functions of $%
\alpha $ and $\varepsilon $. Formula (\ref{form2}) shows that the $\alpha $%
-powers smaller than $-k+1-\widetilde{p}_{k}$, if any, belong only to the
evanescent sector.

The proof is done by induction, studying the pole cancellation in (\ref
{redux}). Assume that (\ref{form2}) holds for $k<n$. A theorem derived in
ref. \cite{renscal} states that the maximal $\varepsilon $-pole of a Feynman
diagram with $V$ vertices and $L$ loops is at most of order equal to $\min
(V-1,L)$. Then the contributions to $\overline{\Delta }_{n}(\alpha ,f(\alpha
,\varepsilon ),\varepsilon )$ from the diagrams $G$ with $n_{k}$ irrelevant
vertices of level $k$, $L$ loops, $v_{4}$ marginal four-leg vertices and $%
V=v_{4}+\sum_{k<n}n_{k}$ vertices have the form 
\begin{eqnarray}
\overline{\Delta }_{n}(\alpha ,f(\alpha ,\varepsilon ),\varepsilon )
&=&\sum_{G}\frac{\varepsilon ^{s}}{\varepsilon ^{\min (V-1,L)}}\alpha
^{L-1}\prod_{k<n}\overline{f}_{k}^{n_{k}}(\alpha ,\varepsilon )  \nonumber \\
&=&\sum_{G}\frac{\varepsilon ^{s^{\prime }}}{\varepsilon ^{\min (V-1,L)}}%
\alpha ^{L-1+t+\sum_{k<n}n_{k}q_{k}}\prod_{k<n}\left( \frac{\varepsilon }{%
\alpha }\right) ^{j_{k}},  \label{ad}
\end{eqnarray}
where $s,s^{\prime },t,j_{k}$ are non-negative integers and $j_{k}\leq
n_{k}q_{k}$ and $\sum_{k<n}kn_{k}=n$. The factor $\alpha ^{L-1}$ comes from
the $\alpha $-powers attached to the vertices, as shown in the previous
sections. The factors $\varepsilon ^{s}$, $\varepsilon ^{s^{\prime }}$ are
inserted in (\ref{ad}) to take care of the subleading divergences and the
extra powers of $\varepsilon $ coming from $\overline{f}_{j,k}(\alpha
,\varepsilon )$, while $\alpha ^{t}$ takes care of the extra powers of $%
\alpha $ coming from $\overline{f}_{j,k}(\alpha ,\varepsilon )$. The
coefficients of the sum are not important for our analysis and so are left
unspecified. Specializing to the simple poles we get contributions of the
type 
\[
\frac{1}{\varepsilon }\alpha ^{L-\min
(V-1,L)+\sum_{k<n}n_{k}q_{k}+t+s^{\prime }} 
\]
Formula (\ref{legs}) can be written as 
\[
v_{4}+\sum_{k<n}n_{k}\widetilde{p}_{k}=\widetilde{p}_{n}+L. 
\]
Using this relation and (\ref{qk}), it is immediate to derive the inequality 
\begin{equation}
L-\min (V-1,L)+\sum_{k<n}n_{k}q_{k}\geq q_{n},  \label{inequa}
\end{equation}
so the $\alpha $-exponent of the simple pole of $\overline{\Delta }_{n}$ is
always $\geq q_{n}$, and the simple poles contained in $\overline{\Delta }%
_{n}(\alpha ,\overline{f}(\alpha ,\xi ,\varepsilon ),\varepsilon )$ are
multiplied by powers $\alpha ^{q_{n}+s}$, $s\geq 0$.

With the ansatz (\ref{anx}), in (\ref{redux}) the coefficient $\overline{%
\zeta }_{n,k}$ is multiplied by a sum of objects of the form 
\begin{equation}
\varepsilon ^{q_{n}+k}\left( \frac{\alpha }{\varepsilon }\right) ^{m}\alpha
^{r},  \label{lead2}
\end{equation}
with $m,r\geq 0$. The simple pole is 
\[
\frac{1}{\varepsilon }\alpha ^{q_{n}+1+k+r}. 
\]
In total, the simple poles of (\ref{redux}) have the form 
\begin{equation}
\frac{\alpha ^{q_{n}+1}}{\varepsilon }\left( \sum_{s\geq 0}a_{s}\alpha
^{s}+\sum_{k,r\geq 0}\overline{\zeta} _{n,k}c_{k,r}\alpha ^{k+r}\right) ,
\label{get}
\end{equation}
where $a_{s}$ and $c_{k,r}$ are known numerical factors. Thus, if the
coefficients of $\overline{\zeta }_{n,j}\alpha ^{j}$ are nonzero it is
possible to determine $\overline{\zeta }_{n,j}$ iteratively in $j$ from the
cancellation of the pole. Finally, using (\ref{llog}) the term $\overline{%
\zeta }_{n,j}\alpha ^{j}$ inside the parenthesis of (\ref{get}) is
multiplied by the coefficient 
\[
\frac{\left( -\beta _{\alpha }^{(1)}\right) ^{q_{n}+j+1}}{(q_{n}+j+1)!}%
\prod_{i=0}^{q_{n}+j}\left( \frac{\gamma _{n\ell }^{(1)}-n\gamma _{\ell
}^{(1)}}{\beta _{\alpha }^{(1)}}-i\right) , 
\]
thus the invertibility conditions are again (\ref{inveps}).

Once the poles have been cancelled out and the constants $\overline{\zeta }%
_{n,k}$ are determined, collecting (\ref{ad})\ and (\ref{lead2}) we obtain 
\begin{equation}
\overline{f}_{n}\sim \alpha ^{q_{n}}\left[ \sum\Sb L\geq 1,\text{~}%
u,s,i_{n}\geq 0  \\  \\ L+i_{n}\leq q_{n}+s^{\prime }+u  \endSb \left( \frac{%
\varepsilon }{\alpha }\right) ^{q_{n}-L-i_{n}}\alpha ^{t}\varepsilon
^{s^{\prime }+u}+\sum\Sb m,r,j\geq 0  \\  \\ m\leq q_{n}+j  \endSb \overline{%
\zeta }_{n,j}\varepsilon ^{j}\alpha ^{r}\left( \frac{\varepsilon }{\alpha }%
\right) ^{q_{n}-m}\right] ,  \label{beha}
\end{equation}
having written $\sum_{k<n}j_{k}=\sum_{k<n}q_{k}n_{k}-i_{n}$, $i_{n}\geq 0$,
and $\sum_{k<n}q_{k}n_{k}-\min (V-1,L)=q_{n}-L+u$, $u\geq 0$ (see (\ref
{inequa})). We see that $\overline{f}_{n}(\alpha ,\varepsilon )$ has the
form (\ref{form2}) for $k=n$, which reproduces the inductive hypothesis.

\bigskip

In the absence of three-leg vertices, the invertibility conditions found in 
\cite{renscal,infredfin} read in the notation of this paper (recall that
here $\gamma _{n\ell }$ denotes the anomalous dimensions of the operators $%
g^{2p_{n}}\mathcal{O}_{n}(\varphi )$) 
\begin{equation}
r_{n}\equiv \tau _{n}+n-1+\widetilde{p}_{n}\notin \mathbb{N}.  \label{inv}
\end{equation}
These invertibility conditions can be more or less restrictive than the
regularized invertibility conditions (\ref{inveps}), but they are not in
contradiction with (\ref{inveps}) in the physical limit. Recall that both (%
\ref{inv}) and (\ref{inveps}) are sufficient, but not necessary conditions
and in particular cases they can be relaxed. If $n-1+\widetilde{p}_{n}>0$
the conditions (\ref{inveps}) are less restrictive than (\ref{inv}). In this
case the violations of (\ref{inv}) that are not violations of (\ref{inveps})
do not cause the introduction of any new parameters. If $n-1+\widetilde{p}%
_{n}<0$ the conditions (\ref{inveps}) are more restrictive than (\ref{inv}).
Then there are situations where (\ref{inveps}) are violated but (\ref{inv})
are fulfilled. This happens if $\tau _{n}=N\in \mathbb{N}$ and $%
r_{n}=N-q_{n}<0$. In this case it is necessary to introduce new independent
parameters, which however affect only the evanescent sector, with no
physical consequences. Indeed, formula (\ref{form2}) for $k=n$ shows that
the physical function $\overline{f}_{n}(\alpha ,0)$ is not interested by the
violation of the invertibility condition, because it starts with the power $%
\alpha ^{q_{n}}$, while the invertibility problem occurs only when the power 
$\alpha ^{N}$ is present. Thus the $\varepsilon =0$ results of \cite
{infredfin,renscal} and the $\varepsilon \neq 0$ results of \cite{renscal}
are fully recovered.

\section{Explicit leading-log solution}

\setcounter{equation}{0}

In this section we solve the infinite reduction at the leading-log level. We
show that the leading-log approximation is sufficient to derive the
invertibility conditions for the existence of the infinite reduction to all
orders.

At the leading-log level, the beta function of the marginal coupling $\alpha 
$%
\begin{equation}
\frac{\mathrm{d}\alpha }{\mathrm{d}\ln \mu }=\widehat{\beta }_{\alpha
}(\alpha ,\varepsilon )=\beta _{\alpha }\left( \alpha \right) -\varepsilon
\alpha =-\varepsilon \alpha +\beta _{1}\alpha ^{2}+\mathcal{O}(\alpha ^{3})
\label{beta3}
\end{equation}
and the anomalous dimensions 
\[
\gamma _{n\ell }(\alpha )=\gamma _{n\ell }^{(1)}\alpha +\mathcal{O}(\alpha
^{2}) 
\]
can be truncated to one loop. The models with and without three-leg marginal
vertices can be treated with a unique formalism, using formula (\ref{form2}%
), where it is understood that, in the presence of three-leg marginal
vertices $q_{k}$ is just equal to zero. The leading-log approximation
amounts to take the lowest $\alpha $-powers at $\varepsilon =0$ and the
corresponding $\alpha $ powers at higher orders in $\varepsilon $. More
precisely, in this approximation $\overline{f}_{k}(\alpha ,\varepsilon )$
and $\overline{\delta }_{n}(\alpha ,\varepsilon )$ read 
\[
\overline{f}_{k}(\alpha ,\varepsilon )=\alpha ^{q_{k}}\sum_{j=0}^{q_{k}}%
\overline{f}_{j,k}\left( \frac{\varepsilon }{\alpha }\right) ^{j},\qquad 
\overline{\delta }_{n}(\alpha ,\varepsilon )=\alpha
^{q_{n}}\sum_{j=0}^{q_{n}}d_{j,n}\left( \frac{\varepsilon }{\alpha }\right)
^{j}, 
\]
where $q_{n}$ is an integer and $\overline{f}_{j,k}$, $d_{j,n}$ are
constants. The solution (\ref{sola}) is 
\begin{eqnarray}
f_{n}(\alpha ,\xi _{n},\varepsilon ) &=&-\frac{\alpha ^{q_{n}}\,}{\beta
_{\alpha }^{(1)}}\sum_{j=0}^{q_{n}}\frac{d_{j,n}\varepsilon ^{j}}{\alpha
^{j}(\tau _{n}-q_{n}+j)}~_{2}F_{1}\left[ 1,j-q_{n},\tau _{n}-q_{n}+j+1,\frac{%
\varepsilon }{\alpha \beta _{\alpha }^{(1)}}\right]  \nonumber \\
&&+\xi _{n}\left( \alpha \beta _{\alpha }^{(1)}-\varepsilon \right) ^{\tau
_{n}}.  \label{renolog}
\end{eqnarray}
Observe that the hypergeometric functions appearing in the sum are
polynomial, since $q_{n}-j$ is a non-negative integer.

At the level of bare couplings, the reduction has the form (\ref{abr}).
Manipulating the formulas given above and using (\ref{limit}), the formula
for $\zeta _{n}(\xi _{n},\varepsilon )$ can be derived. The result is 
\begin{equation}
\zeta _{n}(\xi _{n},\varepsilon )=\varepsilon
^{q_{n}}\sum_{j=0}^{q_{n}}d_{j,n}\left( -\beta _{\alpha }^{(1)}\right)
^{j-q_{n}-1}\frac{\Gamma \left( \tau _{n}-q_{n}+j\right) \Gamma \left(
q_{n}-j+1\right) }{\Gamma \left( \tau _{n}+1\right) }+\xi _{n}(-\varepsilon
)^{\tau _{n}}.  \label{cn}
\end{equation}

We see again that if $\tau _{n}\notin $ $\mathbb{N}$ the relation $\zeta
_{n}(\xi _{n},\varepsilon )$ is not analytic in $\xi _{n}$ and $\varepsilon $%
. Both the bare and renormalized reduction relations are analytic in $%
\varepsilon $ only for $\xi _{n}=0$, which gives 
\begin{equation}
\overline{\zeta }_{n}(\varepsilon )=\varepsilon
^{q_{n}}\sum_{j=0}^{q_{n}}d_{j,n}\left( -\beta _{\alpha }^{(1)}\right)
^{j-q_{n}-1}\frac{\Gamma \left( \tau _{n}-q_{n}+j\right) \Gamma \left(
q_{n}-j+1\right) }{\Gamma \left( \tau _{n}+1\right) }.  \label{er}
\end{equation}
This formula uniquely determines the reduction.

For $q_{n}=0$ we simply have 
\[
f_{n}(\alpha ,\xi _{n},\varepsilon )=-\frac{d_{0,n}}{\tau _{n}\beta _{\alpha
}^{(1)}}~+\xi _{n}\left( \alpha \beta _{\alpha }^{(1)}-\varepsilon \right)
^{\tau _{n}},\qquad \overline{\zeta }_{n}(\varepsilon )=-\frac{d_{0,n}}{\tau
_{n}\beta _{\alpha }^{(1)}}. 
\]

\bigskip

\textbf{Violations of the invertibility conditions.} It is interesting to
describe the appearance of new parameters, when the invertibility conditions
are violated, in the leading-log approximation. Assume that some regularized
invertibility conditions (\ref{inveps}) are violated, i.e. $\tau _{n}=%
\overline{r}\in \mathbb{N}$. To study this situation it is convenient to
approach it continuously from $\tau _{n}=\overline{r}+\delta $ and then take
the limit $\delta \rightarrow 0$. If $\overline{r}>q_{n}$ this limit is
trivial in the leading-log approximation, so we just need to discuss the
case $\overline{r}\leq q_{n}$.

Collecting the singular terms of (\ref{renolog}) we get an expression of the
form 
\[
f_{n}(\alpha ,\xi _{n},\varepsilon )=\left( \alpha \beta _{\alpha
}^{(1)}-\varepsilon \right) ^{\overline{r}}\left\{ \frac{a}{\delta }%
\varepsilon ^{q_{n}-\overline{r}}\,+\xi _{n}\left[ 1+\delta \ln \left(
\alpha \beta _{\alpha }^{(1)}-\varepsilon \right) \right] \right\} +\alpha
^{q_{n}}P_{n}(\varepsilon /\alpha )+\mathcal{O}(\delta ,\xi _{n}\delta
^{2}), 
\]
where $a$ is a known numerical factor and $P_{n}(\varepsilon /\alpha )$ is a
certain $\xi $- and $\delta $-independent polynomial of degree $q_{n}$. The $%
\delta $-singularity can be removed redefining $\xi _{n}$ as 
\[
\xi _{n}=-\frac{a}{\delta }\varepsilon ^{q_{n}-\overline{r}}+\xi
_{n}^{^{\prime }}, 
\]
thus obtaining a non-singular expression 
\[
f_{n}(\alpha ,\xi _{n},\varepsilon )=\left( \alpha \beta _{\alpha
}^{(1)}-\varepsilon \right) ^{\overline{r}}\left\{ \xi _{n}^{^{\prime
}}-a\varepsilon ^{q_{n}-\overline{r}}\ln \left( \alpha \beta _{\alpha
}^{(1)}-\varepsilon \right) \right\} +\alpha ^{q_{n}}P_{n}(\varepsilon
/\alpha ). 
\]
Finally, the relation between the bare and renormalized constants $\zeta
_{n} $ and $\xi _{n}^{\prime }$ at $\zeta _{k}=0$, $k<n$, read

\begin{equation}
\zeta _{n}(\xi ,\varepsilon )=\lim_{\alpha \rightarrow 0}f_{n}(\alpha ,\xi
_{n},\varepsilon )=(-\varepsilon )^{\overline{r}}\left[ \xi _{n}^{^{\prime
}}-a\varepsilon ^{q_{n}-\overline{r}}\ln (-\varepsilon )+b\varepsilon
^{q_{n}-\overline{r}}\right] ,  \label{yt}
\end{equation}
where $b$ is another known numerical factor, originated by $\alpha
^{q_{n}}P_{n}(\varepsilon /\alpha )$. We see that no choice of the constant $%
\xi _{n}^{^{\prime }}$ is able to remove the analyticity violation in both
the bare and renormalized reduction relations. The violation can be hidden
in a new independent coupling, but since $\ln (-\varepsilon )$ is multiplied
by $\varepsilon ^{q_{n}-\overline{r}}$ it is sufficient to write $\xi
_{n}^{^{\prime }}=\varepsilon ^{q_{n}-\overline{r}}\xi _{n}^{^{\prime \prime
}}$ and associate the new coupling with $\xi _{n}^{^{\prime \prime }}$.
Therefore the new coupling belongs to the evanescent sector if $\overline{r}%
<q_{n}$, it is physical if $\overline{r}=q_{n}$.

\section{Irrelevant deformations in the presence of several marginal
couplings}

\setcounter{equation}{0}

In this section we describe the construction of irrelevant deformations when
the renormalizable subsector $\mathcal{R}$ contains more independent
marginal couplings.

Consider a renormalizable theory with two marginal couplings $\alpha $ and $%
\rho =\alpha \eta $. It is convenient, for intermediate purposes, to express 
$\eta $ as a function $\widetilde{\eta }(\alpha ,\xi ,\varepsilon )$ of $%
\alpha $ and an arbitrary constant $\xi $, as explained in section 2,
solving the equations (\ref{zimmesp}). Then the marginal sector is described
by a unique running marginal coupling, $\alpha $, plus an arbitrary
constant, and most of the arguments of the infinite reduction proceed as in
the presence of a single marginal coupling. As before, we state two
equivalent criteria for the infinite reduction. The first criterion is based
on the analyticity properties of the renormalized reduction relations. The
second criterion is based on the comparison between the analyticity
properties of the renormalized and bare reduction relations.

Precisely, when certain invertibility conditions, derived below, hold, the
reduction is uniquely determined by the requirement that

\noindent 1) the renormalized reduction relations be perturbatively
meromorphic in $\alpha $, analytic in $\eta $ and analytic in $\varepsilon $;

\noindent or, equivalently, by the requirement that

\noindent 2) the renormalized reduction relations be analytic in $%
\varepsilon $ and $\xi $ and at the same time the bare relations be analytic
in $\varepsilon $ and $\xi (-\varepsilon )^{Q}$, where $Q$ is a non-integer
exponent that can possibly depend on $\varepsilon $.

The analyticity requirements in $\xi $ and $\xi (-\varepsilon )^{Q}$ are
obviously due to the presence of the second marginal coupling.

We focus on the leading-log approximation, for simplicity, whose beta
functions (\ref{betas}) have the one-loop coefficients (\ref{betaso}). For
definiteness, we choose the positive sign in front of $s$ in the leading-log
solution (\ref{renoz}). We study the $2\ell $-level terms belonging to the
queue of the irrelevant deformation, using the minimal parametrization 
\[
\lambda _{\ell }\mathcal{O}_{\ell }(\varphi )+\lambda _{2\ell }\mathcal{O}%
_{2\ell }(\varphi )+\cdots , 
\]
and assuming that the beta functions of the irrelevant couplings are $%
\widehat{\beta }_{\ell }=\beta _{\ell }-\varepsilon \lambda _{\ell }$, $%
\widehat{\beta }_{2\ell }=\beta _{2\ell }-\varepsilon \lambda _{2\ell }$, in
the minimal subtraction scheme. For $\alpha $ small the lowest-order beta
functions of $\lambda _{\ell }$ and $\lambda _{2\ell }$ have generically the
forms 
\begin{equation}
\beta _{\ell }=\lambda _{\ell }\alpha (d+e\eta ),\qquad \beta _{2\ell
}=\lambda _{2\ell }\alpha (f+g\eta )+h\lambda _{\ell }^{2},  \label{duebeta}
\end{equation}
where $d,e,f,g,h$ are numerical factors. The coupling $\lambda _{2\ell }$ is
related to $\lambda _{\ell }$ and $\alpha ,\eta $ by a relation of the form 
\begin{equation}
\lambda _{2\ell }=f_{2}(\alpha ,\eta ,\varepsilon )\lambda _{\ell }^{2}.
\label{ry}
\end{equation}
The function $f_{2}(\alpha ,\eta ,\varepsilon )$ can be worked out using a
procedure similar to the one described in appendix A for multivariable
renormalization constants. Define the function 
\begin{equation}
\widetilde{f}_{2}(\alpha ,\xi ,\varepsilon )=f_{2}\left( \alpha ,\widetilde{%
\eta }(\alpha ,\xi ,\varepsilon ),\varepsilon \right) .  \label{Ftilde}
\end{equation}
Differentiating (\ref{ry}) and using (\ref{duebeta}), we find the equation 
\begin{equation}
\widehat{\beta }_{\alpha }\left( \alpha ,\widetilde{\eta }(\alpha ,\xi
,\varepsilon ),\varepsilon \right) \ \frac{\mathrm{d}\widetilde{f}%
_{2}(\alpha ,\xi ,\varepsilon )}{\mathrm{d}\alpha }+2\alpha \widetilde{f}%
_{2}(\alpha ,\xi ,\varepsilon )\left( \widetilde{d}+\widetilde{e}\widetilde{%
\eta }(\alpha ,\xi ,\varepsilon )\right) -\varepsilon \widetilde{f}%
_{2}(\alpha ,\xi ,\varepsilon )=h,  \label{RGFtilde}
\end{equation}
where $\widetilde{d}=d-f/2$ and $\widetilde{e}=e-g/2$. The solution depends
on $\xi $ and a further arbitrary constant $k_{2}$. Eliminating $\xi $ with
the help of (\ref{c1}), the solution reads 
\begin{equation}
f_{2}(\alpha ,\eta ,\varepsilon ,k_{2})=\overline{f}_{2}(\alpha ,\eta
)+k_{2}~s_{2}(\alpha ,\eta ,\varepsilon ),  \label{solr2}
\end{equation}
where 
\begin{equation}
\overline{f}_{2}(\alpha ,\eta )=\frac{h(1-z)}{\alpha s(\gamma -1)}%
~_{2}F_{1}[1,\gamma -2\widetilde{e}/c,\gamma ,z],\qquad s_{2}(\alpha ,\eta
,\varepsilon )=\frac{1}{\alpha }z^{1-\gamma }(1-z)^{2\widetilde{e}/c},
\label{r2bar}
\end{equation}
with 
\[
\gamma =1+\frac{\widetilde{e}}{c}+\frac{1}{s}\left( 2\widetilde{d}-\beta
_{1}-b\frac{\widetilde{e}}{c}\right) , 
\]
and $z=\xi (\beta _{1}\alpha -\varepsilon )^{s/\beta _{1}}$. The quantity $%
k_{2}$ is constant along the RG flow, and can be viewed as a function of $%
\xi $ and $\varepsilon $.

The analyticity properties of the solution can be analyzed in the $\alpha $-$%
\xi $ parametrization. The function $\widetilde{\eta }$ of (\ref{renoz}) is
analytic in $\alpha $, $\varepsilon $ and $z=\xi (\beta _{1}\alpha
-\varepsilon )^{s/\beta _{1}}$. The special solution $\overline{f}%
_{2}(\alpha ,\eta )$, on the other hand, is meromorphic in $\alpha $,
analytic in $\varepsilon $ and analytic in $z$ at $z=0$, so it satisfies the
requirement 1) stated above.

To study the arbitrariness of the general solution $f_{2}(\alpha ,\eta
,\varepsilon )$ write, for example, $k_{2}(\xi ,\varepsilon )=k_{2}^{\prime
}\xi ^{p}\varepsilon ^{q}$. If 
\begin{equation}
(1-\gamma )\frac{s}{\beta _{1}}=m+n\frac{s}{\beta _{1}},\qquad m,n\in %
\mathbb{N},  \label{mn}
\end{equation}
then taking $p=n-1+\gamma $ and $q\in \mathbb{N}$ we have 
\[
k_{2}(\xi )\ s_{2}(\alpha ,\eta ,\varepsilon )=k_{2}^{\prime }\frac{1}{%
\alpha }\varepsilon ^{q}(\beta _{1}\alpha -\varepsilon )^{m}\left( \xi
(\beta _{1}\alpha -\varepsilon )^{s/\beta _{1}}\right) ^{n}(1-z)^{2%
\widetilde{e}/c}, 
\]
which is meromorphic in $\alpha $ with the right behavior at $\alpha \sim 0$%
, analytic in $\varepsilon $ and in $\xi (\beta _{1}\alpha -\varepsilon
)^{s/\beta _{1}}$. Thus the invertibility conditions for requirement 1) are
that there should exist no pair of integers $m$, $n$ such that (\ref{mn})
holds. Assuming that $s$ is irrational or complex, which happens in most
cases, and recalling that the ratios of one-loop coefficients are rational
numbers, the invertibility conditions are equivalent to 
\begin{equation}
-\frac{\widetilde{e}}{c}\notin \mathbb{N}\text{\qquad or}\qquad 1+\frac{b%
\widetilde{e}}{c\beta _{1}}-\frac{2\widetilde{d}}{\beta _{1}}\notin %
\mathbb{N}.  \label{iden}
\end{equation}
It is sufficient to fulfill one of the two conditions (\ref{iden}) to fix $%
k_{2}^{\prime }=0$ and uniquely determine the reduction by requirement 1).

The special solution $\overline{f}_{2}(\alpha ,\eta )$ is analytic in $z=0$.
Using the $k_{2}$-freedom it is possible to have special solutions that are
analytic in $z=1$ or $z=\infty $ \cite{infredfin}. The existence conditions
around $z=\infty $ are 
\[
\left( 1-\gamma +2\frac{\widetilde{e}}{c}\right) \frac{s}{\beta _{1}}\neq m-n%
\frac{s}{\beta _{1}},\qquad m,n\in \mathbb{N}, 
\]
which are equivalent to (\ref{iden}) if $s$ is irrational or complex.
Finally, around $z=1$ the invertibility conditions are just 
\begin{equation}
2\frac{\widetilde{e}}{c}-1\notin \mathbb{N}.  \label{aiut}
\end{equation}

\bigskip

Now, let us study the reduction at the bare level. The bare reduction
relations have the form 
\begin{equation}
\lambda _{2\ell \mathrm{B}}=\zeta _{2}(\eta _{\mathrm{B}},\varepsilon )\frac{%
\lambda _{\ell \mathrm{B}}^{2}}{\alpha _{\mathrm{B}}}.  \label{bag}
\end{equation}
The powers of $\lambda _{\ell \mathrm{B}}$ and $\alpha _{\mathrm{B}}$ are
fixed matching the dimensionalities at $\varepsilon \neq 0$ and demanding
analyticity or meromorphy. The relation contains an arbitrary function of
the dimensionless bare coupling $\eta _{\mathrm{B}}$.

Rewrite (\ref{bag}) in terms of the renormalized couplings, 
\begin{equation}
f_{2}Z_{2\ell }=\frac{\zeta _{2}Z_{\ell }^{2}}{\alpha Z_{\alpha }},
\label{zl2}
\end{equation}
then use (\ref{renoz}) to express this equality in terms of $\alpha $, $\xi $%
, and $\varepsilon $. Taking the limit $\alpha \rightarrow 0$, where all $Z$%
s tend to one, the formula 
\[
\zeta _{2}=\lim_{\alpha \rightarrow 0}\alpha \widetilde{f}_{2}(\alpha ,\xi
,\varepsilon ,k_{2}) 
\]
is obtained, which allows us to compute the constant $\zeta _{2}$ as a
function of $\xi $, $k_{2}$ and $\varepsilon $. The result is 
\begin{equation}
\zeta _{2}=\frac{h(1-z_{\mathrm{B}})}{s(\gamma -1)}~_{2}F_{1}[1,\gamma -2%
\widetilde{e}/c,\gamma ,z_{\mathrm{B}}]+k_{2}\ z_{\mathrm{B}}^{1-\gamma
}(1-z_{\mathrm{B}})^{2\widetilde{e}/c},  \label{ery}
\end{equation}
where $z_{\mathrm{B}}=\xi (-\varepsilon )^{s/\beta _{1}}$.

Now we show that the reduction is uniquely fixed also demanding that the
renormalized relations be analytic in $\varepsilon $ and $\xi $ and, at the
same time, the bare relations be analytic in $\varepsilon $ and $\xi
(-\varepsilon )^{s/\beta _{1}}$, without paying attention to the $\alpha $
-dependence. We have shown in section 2 that at $\alpha \neq 0$ the function 
$\widetilde{\eta }(\alpha ,\xi ,\varepsilon )$ is analytic in $\xi $ and $%
\varepsilon $. With arguments similar to the ones of section 2 it is
possible to define the arbitrary constant $k_{2}$ of $\widetilde{f}%
_{2}(\alpha ,\xi ,\varepsilon ,k_{2})\equiv f_{2}(\alpha ,\widetilde{\eta }%
(\alpha ,\xi ,\varepsilon ),\varepsilon ,k_{2})$ in such a way that $%
\widetilde{f}_{2}$ is analytic in $\varepsilon $ and $\xi $. In formula (\ref
{solr2}) this can be achieved writing $k_{2}=k_{2}^{\prime }\xi ^{n-1+\gamma
}\varepsilon ^{q}$ with $n,q\in \mathbb{N}$. However, if $q>0$ the
arbitrariness affects only the evanescent sector, with no observable
consequence, so we can take $q=0$. Then (\ref{ery}) immediately shows that
the bare relations are analytic in $\varepsilon $ and $\xi (-\varepsilon
)^{s/\beta _{1}}$ precisely when (\ref{mn}) holds, so the invertibility
conditions are again (\ref{iden}). We conclude that the criterion 2) is
equivalent to the criterion 1).

The same invertibility conditions are found demanding that the renormalized
relations be analytic in $\varepsilon $ and $\xi $, and the bare relations
be analytic in $\varepsilon $ and $\xi (-\varepsilon )^{-s/\beta _{1}}$.
Finally, the reduction is uniquely fixed also demanding that the
renormalized relations be analytic in $\varepsilon $ and $\xi $, and the
bare relations be analytic in $\varepsilon $ and $1-\xi (-\varepsilon
)^{s/\beta _{1}}$, in which case the invertibility conditions are (\ref{aiut}%
).

Beyond the leading-log approximation, the $\alpha $-$\eta $ parametrization
can be used in connection with formula (\ref{anxi}). Alternatively, it is
convenient to use the variables $u^{\prime }$ and $v$ defined in section 2,
and related quantities. Then for example analyticity in $\xi (-\varepsilon
)^{s/\beta _{1}}$ is replaced by analyticity in $\xi (-\varepsilon )^{Q}$,
where $Q$ is defined in equation (\ref{compa}), and so on.

The generalization to theories with more marginal couplings follows the same
guidelights and is left to the reader.

\section{Conclusions}

\setcounter{equation}{0}

The field-theoretical investigation of non-renormalizable interactions can
clarify some aspects of quantum field theory that have so far been
underestimated and allows us to explore new ideas for quantum gravity and,
more generally, physics beyond the Standard Model. At the conceptual level,
there is no strict physical reason why a theory should be discarded just
because it is not power-counting renormalizable. In this paper and related
ones we have shown that there is not even a \textit{practical} reason to
justify its exclusion: in a wide class of non-renormalizable theories
calculations are doable in a perturbative fashion and the number of
independent couplings can be kept finite consistently with renormalization.

On the other hand, we cannot forget that power-counting renormalizability
has been the main guidelight to build the Standard Model. Vector bosons and
the Higgs field have been introduced to cure the non-renormalizability of
the Fermi theory of weak interactions. The vector bosons have already been
seen, while the Higgs field will hopefully be discovered at LHC. In view of
these successes, it is hard to deny an important role to power-counting
renormalizability in quantum field theory. Nevertheless, it is mandatory to
understand such a role more precisely. Presumably, power-counting
renormalizability is a preliminary, imperfect version of a deeper selection
principle still well hidden in quantum field theory. The final version of
this principle should leave room also for quantum gravity and new physics
beyond the Standard Model. We hope that our investigations can help
uncovering the final form of the selection principle.

In this paper we have studied the reduction of couplings for renormalizable
and non-re\-nor\-ma\-li\-za\-ble theories at the regularized level. The
dimensional-regularization technique is the most convenient framework to
prove all-order theorems and have control on infinitely many lagrangian
terms. It is possible to formulate the rules of the infinite reduction
comparing the reduction relations at the renormalized and bare levels. If
suitable invertibility conditions are fulfilled, the infinite reduction is
uniquely determined by the contemporary $\varepsilon $-analyticity of the
renormalized and bare relations. When the invertibility conditions are
violated, new couplings appear along the way. It is possible to count the
parameters of the non-renormalizable interaction, or study their
distribution and density, just from knowledge of the renormalizable
subsector $\mathcal{R}$, before turning the non-renormalizable interaction
on.

We have mainly worked with theories where $\mathcal{R}$ contains a unique
marginal coupling, but with some additional effort the results can be
generalized to theories containing more marginal couplings. The marginal
sector has to be fully interacting and it is not possible to switch the
marginal interactions off keeping the irrelevant interaction on.
Generalizations to theories with a free or partially interacting marginal
sector demand further insight, in view of applications to quantum gravity in
four dimensions.

\vskip 25truept \noindent {\Large \textbf{Appendix\quad Derivation of the
renormalization constants from the beta functions}}

\vskip 15truept

\renewcommand{\theequation}{A.\arabic{equation}} \setcounter{equation}{0}

In this appendix I\ describe a general procedure to derive multivariable
renormalization constants from their beta functions, which is useful for
several arguments of the paper.

\medskip

In the minimal subtraction scheme the renormalization constants $Z_{\alpha }$
satisfy RG equations of the form 
\begin{equation}
\frac{\mathrm{d\ln }Z_{\alpha }}{\mathrm{d}\ln \mu }=-\frac{\beta _{\alpha }%
}{\alpha }.  \label{orda}
\end{equation}
When the theory has only one coupling $\alpha $, it is possible to write $%
Z_{\alpha }=Z_{\alpha }(\alpha ,\varepsilon )$ and immediately integrate the
RG\ equation 
\begin{equation}
\frac{\mathrm{d\ln }Z_{\alpha }}{\mathrm{d}\alpha }=-\frac{\beta _{\alpha
}(\alpha )}{\alpha \widehat{\beta }_{\alpha }(\alpha ,\varepsilon )},\qquad
\qquad Z_{\alpha }=\exp \left( -\int_{0}^{\alpha }\frac{\beta _{\alpha
}(\alpha ^{\prime })\mathrm{d}\alpha ^{\prime }}{\alpha ^{\prime }\widehat{%
\beta }_{\alpha }(\alpha ^{\prime },\varepsilon )}\right) .  \label{zl}
\end{equation}
The integral in (\ref{zl}) is well-defined at $\varepsilon \neq 0$.

More generally, let $\alpha ,\lambda _{1},\cdots \lambda _{n}$ denote the
couplings in the minimal parametrization. The problem we want to solve is to
reconstruct the renormalization constants $Z_{\alpha }(\alpha ,\lambda
_{i},\varepsilon )$ and $Z_{i}(\alpha ,\lambda _{i},\varepsilon )$ from the
beta functions of $\alpha $ and $\lambda _{i}$.

The RG equations have the form 
\begin{equation}
\frac{\mathrm{d}\alpha }{\mathrm{d}\ln \mu }=\widehat{\beta }_{\alpha
}(\alpha ,\lambda _{i},\varepsilon )=\beta _{\alpha }(\alpha ,\lambda
_{i})-p\varepsilon \alpha ,\qquad \frac{\mathrm{d}\lambda _{i}}{\mathrm{d}%
\ln \mu }=\widehat{\beta }_{i}(\alpha ,\lambda _{i},\varepsilon )=\beta
_{i}(\alpha ,\lambda _{i})-p_{i}\varepsilon \lambda _{i},  \label{rg}
\end{equation}
where $p$, $p_{i}$ are related to the numbers $N$, $N_{i}$ of legs of the
respective vertices by formula (\ref{piennet}): $p=N/2-1$, $p_{i}=N_{i}/2-1$%
. The bare couplings are 
\[
\alpha _{\mathrm{B}}=\mu ^{p\varepsilon }\alpha Z_{\alpha }(\alpha ,\lambda
_{i},\varepsilon ),\qquad \lambda _{i\mathrm{B}}=\mu ^{p_{i}\varepsilon
}\lambda _{i}Z_{i}(\alpha ,\lambda _{i},\varepsilon ). 
\]
Since a non-trivial theory contains at least a vertex with three legs or
more, we can assume that $p$ is strictly positive. Instead of $\alpha $, it
is convenient to use the variable $\alpha ^{\prime }=\alpha ^{1/p}$, which
has $p^{\prime }=1$. We then suppress the primes on $\alpha $ and $p$, which
is equivalent to assume that $\alpha $ is defined so that it has $p=1$. We
are not assuming that the marginal sector contains a single marginal
coupling, nor that the theory has an interacting marginal sector. It is
convenient, but not necessary, to assume that $\alpha $ is a marginal
coupling.

Next, it is convenient to define non-minimal couplings $\eta _{i}$ such that 
\begin{equation}
\lambda _{i}=\alpha ^{p_{i}}\eta _{i}\text{,\qquad }\frac{\mathrm{d}\eta _{i}%
}{\mathrm{d}\ln \mu }=\overline{\beta }_{i}(\alpha ,\eta _{i}),\qquad \frac{%
\mathrm{d}\alpha }{\mathrm{d}\ln \mu }=\overline{\beta }_{\alpha }(\alpha
,\eta _{i})-\varepsilon \alpha .  \label{arena}
\end{equation}
Write their bare couplings as 
\begin{equation}
\alpha _{\mathrm{B}}=\mu ^{\varepsilon }\alpha \overline{Z}_{\alpha }(\alpha
,\eta _{i},\varepsilon ),\qquad \eta _{i\mathrm{B}}=\eta _{i}+\alpha \Delta
_{i}(\alpha ,\eta _{i},\varepsilon ).  \label{ubbo}
\end{equation}

Diagrammatic arguments analogous to those of sections 2 and 3 (see formulas (%
\ref{ah}) and (\ref{legs})), allow us to prove that $\overline{Z}_{\alpha
}(\alpha ,\eta _{i},\varepsilon )-1$, $\overline{\beta }_{i}(\alpha ,\eta
_{i})$ and $\alpha \Delta _{i}(\alpha ,\eta _{i},\varepsilon )$ are analytic
in $\alpha $ and their $L$-loop contributions are of order $\alpha ^{L}$. A
quicker argument proceeds as follows.

It is sufficient to consider the bare Feynman diagrams, since the
renormalized diagrams inherit their $\alpha $-structure from the bare ones.
By construction, in the non-minimal parametrization (\ref{arena}) $\alpha _{%
\mathrm{B}}$ is the unique bare coupling that has a non-vanishing
dimensionality-defect, equal to one. Now, each loop carries a momentum
integral 
\[
\int \frac{\mathrm{d}^{d}p}{(2\pi )^{d}}, 
\]
where $d$ is the continued spacetime dimension. That means that each loop
carries a dimensionality-defect equal to $-1$, which has to be compensated
by a factor $\alpha _{\mathrm{B}}$. Therefore a power $\alpha _{\mathrm{B}%
}^{L}$ is associated with each $L$-loop integral, as claimed.

Note that the $L$-loop contributions to $\overline{\beta }_{\alpha }(\alpha
,\eta _{i})$ are of order $\alpha ^{L+1}$.

Define the functions $\widetilde{\eta }_{i}(\alpha ,\xi ,\varepsilon )$ as
the solutions of the differential equations 
\begin{equation}
\frac{\mathrm{d}\widetilde{\eta }_{i}}{\mathrm{d}\alpha }=\frac{\overline{%
\beta }_{i}(\alpha ,\widetilde{\eta }_{i})}{\overline{\beta }_{\alpha
}(\alpha ,\widetilde{\eta }_{i})-\varepsilon \alpha },  \label{zima}
\end{equation}
$\xi _{i}$ denoting the arbitrary constants, or, equivalently, as the
solutions of the algebraic bare reduction relations 
\begin{equation}
\eta _{i}+\alpha \Delta _{i}(\alpha ,\eta _{i},\varepsilon )=\zeta _{i},
\label{zimb}
\end{equation}
obtained setting $\eta _{i\mathrm{B}}=$constant, where the $\zeta _{i}$'s
are appropriate functions of the $\xi $'s and $\varepsilon $. Equations (\ref
{zimb}) show that the functions $\widetilde{\eta }_{i}(\alpha ,\xi
,\varepsilon )$ are analytic in $\alpha $ at $\varepsilon \neq 0$. The $%
\alpha $-structures of $\overline{\beta }_{\alpha }(\alpha ,\eta _{i})$ and $%
\overline{\beta }_{i}(\alpha ,\eta _{i})$ allow us to draw the same
conclusion directly from (\ref{zima}).

Similarly, define 
\[
\widetilde{Z}_{\alpha }(\alpha ,\xi ,\varepsilon )\equiv \overline{Z}%
_{\alpha }(\alpha ,\widetilde{\eta }_{i}(\alpha ,\xi ,\varepsilon
),\varepsilon ),\qquad \widetilde{\Delta }_{i}(\alpha ,\xi ,\varepsilon
)\equiv \Delta _{i}(\alpha ,\widetilde{\eta }_{i}(\alpha ,\xi ,\varepsilon
),\varepsilon ). 
\]
Written in this form, the renormalization constants satisfy ordinary
first-order differential equations, obtained differentiating (\ref{ubbo}): 
\begin{equation}
\frac{\mathrm{d\ln }\widetilde{Z}_{\alpha }}{\mathrm{d}\alpha }=-\frac{1}{%
\alpha }\frac{\overline{\beta }_{\alpha }(\alpha ,\widetilde{\eta }_{i})}{%
\beta _{\alpha }(\alpha ,\widetilde{\eta }_{i})-\varepsilon \alpha },\qquad 
\frac{\mathrm{d}\left( \alpha \widetilde{\Delta }_{i}\right) }{\mathrm{d}%
\alpha }=-\frac{\overline{\beta }_{i}(\alpha ,\widetilde{\eta }_{i})}{%
\overline{\beta }_{\alpha }(\alpha ,\widetilde{\eta }_{i})-\varepsilon
\alpha }.  \label{orda4}
\end{equation}
Integrate these equations, with the initial conditions 
\begin{equation}
\widetilde{Z}_{\alpha }(0,\xi ,\varepsilon )=1,\qquad \qquad \widetilde{%
\Delta }_{i}(0,\xi ,\varepsilon )<\infty .  \label{incon}
\end{equation}
Such initial conditions are ensured by the $\alpha $-structure proved above
and the $\alpha $-analyticity of $\widetilde{\eta }_{i}(\alpha ,\xi
,\varepsilon )$ at $\varepsilon \neq 0$. Observe that the solutions $\alpha 
\widetilde{\Delta }_{i}$ are immediate to find: 
\[
\alpha \widetilde{\Delta }_{i}(\alpha ,\xi ,\varepsilon )=\widetilde{\eta }%
_{i}(0,\xi ,\varepsilon )-\widetilde{\eta }_{i}(\alpha ,\xi ,\varepsilon ). 
\]
Inverting the functions $\widetilde{\eta }_{i}(\alpha ,\xi ,\varepsilon )$,
write the $\xi _{i}$'s as RG-invariant functions of the couplings and $%
\varepsilon $: 
\begin{equation}
\xi _{i}=\xi _{i}(\alpha ,\eta _{j},\varepsilon ).  \label{invfuncto}
\end{equation}
Next, inserting (\ref{invfuncto}) into $\widetilde{Z}_{\alpha }(\alpha ,\xi
,\varepsilon )$ and $\alpha \widetilde{\Delta }_{i}(\alpha ,\xi ,\varepsilon
)$, $\overline{Z}_{\alpha }(\alpha ,\eta _{i},\varepsilon )$ and $\alpha
\Delta _{i}(\alpha ,\eta _{i},\varepsilon )$ are obtained. Finally, using $%
\eta _{i}=\alpha ^{-p_{i}}\lambda _{i}$ the renormalization constants 
\[
Z_{\alpha }(\alpha ,\lambda _{i},\varepsilon ),\qquad Z_{i}(\alpha ,\lambda
_{i},\varepsilon ). 
\]
are successfully reconstructed.

\end{document}